\documentclass[fleqn,usenatbib]{mnras}

\usepackage{latexsym,mathrsfs,amssymb,bm}
\usepackage[tbtags]{amsmath}
\usepackage[T1]{fontenc}
\usepackage{ae,aecompl,times}
\usepackage{epsfig, ulem}
\usepackage[dvipsnames]{xcolor}
\usepackage{multirow}
\usepackage{tikz,array,color,float}
\usepackage[utf8]{inputenc}
\usepackage[T1]{fontenc}
\usepackage{graphicx}
\usepackage{caption}
\usepackage{multirow}
\usepackage{pdfpages}
\usepackage{adjustbox}

\newcommand{\Mpch}{\,h^{-1}{\rm Mpc}}

\newcommand{\Msh}{\,h^{-1}{\rm M}_\odot}

\newcommand{\Msol}{\,{\rm M}_\odot}

\title[WL and IA with MTNG]{Fully non-linear simulations of galaxy intrinsic alignments for weak lensing with the MillenniumTNG lightcone}

\author[F. Ferlito et al.]{%
\parbox{0.98\textwidth}{
Fulvio Ferlito$^{1}$\thanks{E-mail: \href{mailto:ferlito@mpa-garching.mpg.de}{ferlito@mpa-garching.mpg.de}},
Volker Springel$^{1}$,
Christopher T. Davies$^{2}$,
Toshiki Kurita$^{1}$,
Ana Maria Delgado$^{3}$  \\
Sownak Bose$^{4}$,
and Lars Hernquist$^{5}$
}
\vspace*{0.2cm}\\%
$^{1}$Max-Planck-Institut f\"ur Astrophysik, Karl-Schwarzschild-Str. 1, D-85748, Garching, Germany\\%
$^{2}$Faculty of Physics, Ludwig-Maximilians-Universit\"at, Scheinerstr. 1, 81679 Munich, Germany\\%
$^{3}$Department of Physics \& Astronomy, Johns Hopkins University, 3400 N. Charles Street, Baltimore, MD 21218, USA\\%
$^{4}$Institute for Computational Cosmology, Department of Physics, Durham University, South Road, Durham, DH1 3LE, UK\\%
$^{5}$Center for Astrophysics | Harvard $\&$ Smithsonian, 60 Garden St, Cambridge, MA 02138, USA\\%
}

\date{Accepted XXX. Received YYY; in original form ZZZ}
\pubyear{2025}

\begin{document}

\label{firstpage}

\pagerange{\pageref{firstpage}--\pageref{lastpage}}

\maketitle

\begin{abstract}
We present a complete forward model of a realistic weak lensing galaxy catalogue based on the $740\,\textrm{Mpc}$ hydrodynamical \textsc{MillenniumTNG} (MTNG) simulation. Starting with a complete particle and cell lightcone covering one octant of the sky with redshift range $0<z<1.5$, we apply a group and subhalo finder to generate the corresponding galaxy catalogue for a fiducial observer. For all galaxies,   we compute both their intrinsic and lensing-induced shear. The intrinsic component is derived from the luminosity-weighted inertia tensor of stellar particles, while the extrinsic (gravitational) shear is obtained through full-sky ray-tracing on the same lightcone. This allows us to directly predict the impact of intrinsic alignment (IA) of galaxies on the shear correlation function and popular convergence statistics in a fully non-linear forward model. We find that IA modifies the convergence power spectrum at all angular scales by up to 20\%, it significantly impacts the PDF, altering its tails by 10–20\%, and distorts peak and minimum counts up to 30\%, depending on redshift and scale. We also evaluate the impact of the IA signal on the shear correlation function finding that, along with a redshift dependence, the signal strongly increases for higher galaxy stellar mass cuts applied to the catalogue. Notably, with the highest stellar mass cut we apply, the intrinsic shear autocorrelation can become comparable to the gravitational shear component on small angular scales. Our results highlight the importance of accurately modeling IA for precision weak lensing cosmology with upcoming Stage IV surveys.
\end{abstract}


\section{Introduction}
\label{sec:intro}

Weak gravitational lensing \citep[WL, for reviews see e.g.][]{Bartelmann2001, Hoekstra2008, Kilbinger2015, Mandelbaum2018, Prat2025} refers to the distortion of the images of distant galaxies caused by the gravitationally induced deflection of light as it travels through the intervening matter field. This phenomenon has already proven to be an essential tool for cosmological inference, particularly with Stage-III surveys conducted in recent years, including KiDS \citep{Hildebrandt2016, Heymans2021, Wright2024}, DES \citep{Abbott2022, Bechtol2025}, and HSC \citep{Aihara2022}. Looking ahead, Stage-IV WL surveys such as Euclid \citep{Amendola_2018, Aussel2025}, Rubin \citep{LSST}, and Roman \citep{Roman} will offer higher angular resolution and broader sky coverage, significantly enhancing the richness of WL datasets and promising more precise measurements. To fully exploit the scientific potential of these surveys, it is crucial to provide highly accurate theoretical predictions. This becomes increasingly challenging as observations probe the non-linear regime, where analytic frameworks break down and physical systematic uncertainties become significant.

In particular, to robustly constrain cosmological parameters with high precision and accuracy, it is crucial to fully understand each of the broad range of systematics which impact WL inference at different levels and in different ways. A non-exhaustive list of the main known physical systematics includes: uncertainties in baryonic physics \citep{White2004,Jing2006,Coulton2020,osato2021,Ferlito2023}, massive neutrinos \citep{Kitching2008,Liu2018,Fong2019}, source clustering \citep{Bernardeau1998,Gatti2024}, and, last but not least, the intrinsic alignment of galaxies \citep[IA, for reviews, see e.g.,][]{Joachimi2015,Kirk2015,Kiessling2015,Lamman2024}, which is the focus of the work presented here.

It is well known that galaxies tend to align with the surrounding matter field \citep{Valdes1983,Miralda1991,Hirata2007}, referred to as galaxy intrinsic alignment (IA). Such alignment, aside from being used as a cosmological probe \citep[see e.g.][]{Chisari2013, Schmidt2015, Akitsu2021, Kurita2023, Schmidt2012b, Akitsu2023, Taruya2020, Okumura2023, Xu2023}, directly contaminates the WL signal, by introducing an additional correlation between galaxy shapes and alignments that is independent of WL, where the WL signal itself is extracted from galaxy shape measurements \citep{Joachimi2010,Zhang2010,Troxel2012}. 

The degree to which IA contaminates WL measurements is still not well constrained, and efforts to quantify and describe the IA signal have resulted in a range of analytical models that span different approaches and degrees of sophistication. Examples include the \textit{Linear Alignment} model \citep[LA,][]{Catelan2001, Hirata2004}, \textit{Nonlinear Alignment} model \citep[NLA,][]{Bridle2007, Joachimi2011}, \textit{Tidal Alignment and Tidal Torquing} model \citep[TATT,][]{Blazek2019}, hybrid Lagrangian models \citep[e.g. HYMALAYA,][]{Maion2024}, effective field theory models \citep[e.g.][]{Vlah2020, Vlah2021, Bakx2023, Chen2024}, and models based on the halo model formalism \citep{Schneider2010,Fortuna2021}. These analytical models can only be incorporated into summary statistics for which analytic models also exist, such as the WL power spectrum. It is generally nontrivial to extend these analytic models to higher-order statistics, where one instead needs to rely on measurements from simulations. Recent strategies to address this problem include the \textit{}{IA infusion} approach \citep{Harnois2022}, which uses the same mass shells employed in numerical WL ray-tracing to forward model the IA signature.

The intrinsic alignment of galaxies is a complex phenomenon, which is influenced by a variety of factors ranging from cosmological scales down to sub-galactic scales \citep[see e.g.][]{Troxel2015}, which are highly non linear. For this reason, hydrodynamical simulations have emerged as an essential tool to quantify this effect \citep[a non-exhaustive list includes][]{Chisari2015,Tenneti2015,Velliscig2015b,Hilbert2017,Bate2020,Bhowmick2020,Shi2021,Kurita2021,Samuroff2021,Zjupa2022,Lee2025}. These studies demonstrate that hydrodynamical simulations can meaningfully capture IA, as they incorporate detailed galaxy formation physics and do not assume any relationship between galaxies and the surrounding gravitational field or their host halos. 

Recently, \citet[][hereafter D23]{Delgado2023} used a very large hydrodynamical simulation of galaxy formation, the flagship simulation of the MillenniumTNG (MTNG) project, to measure the projected correlation function of the intrinsic shear of galaxies, and detected IA with the density field at high significance. D23 further detected a significant IA signal for elliptical galaxies assuming the NLA model, a weak IA signal for spirals assuming the TATT model, and a mass dependent misalignment between central galaxies and their host dark-matter halos.

Our work significantly extends the analysis conducted by D23 for the MTNG simulation, with a focus on the lightcone output, to produce a realistic shear catalogue which directly captures both the intrinsic shear, as well as the extrinsic (lensing-induced) shear; i.e.~the pure WL signal. Such a catalogue, which covers an octant of the sky and spans the redshift range  $z=[0, 1.5]$, allows us to directly predict the impact of intrinsic alignments on a range of commonly used WL lensing statistics, in the fully non-linear regime, and without relying on analytic models. The statistics measured here include: shear correlation function, convergence power spectrum, convergence PDF, convergence peaks and minima.

This paper is organized as follows. In Section~\ref{sec:theory}, we provide an overview of the theoretical background of weak lensing (Sec.~\ref{subsec:wl}) and intrinsic alignment (Sec.~\ref{subsec:ellipticity}), then focusing on the correlation function and power spectrum (Sec.~\ref{subsec:ps_corrfunc}). In Section~\ref{sec:methods}, after describing the MillenniumTNG simulations (Sec.~\ref{subsec:mtng}), and the methodology used to construct the galaxy catalogue (Sec.~\ref{subsec:gal_cat_generation}), we show our galaxy selection (Sec.~\ref{subsec:gal_cat_selection}), and describe how we compute intrinsic and extrinsic shear (Sec.~\ref{subsec:intr_shear} and ~\ref{subsec:wl_shear}), and generate convergence maps (Sec.~\ref{subsec:gamma_to_kappa}). In Section~\ref{sec:results}, we present our main results. Starting with a comparison to theoretical prediction (Sec.~\ref{subsec:comp_theory}), we then investigate the redshift dependence (Sec.~\ref{subsec:redshift_dep}) and the mass dependence (Sec.~\ref{subsec:mass_dependence}) of IA on WL. We conclude in Section~\ref{sec:conclusions} with a summary of our findings and an outlook on future applications.


\section{Theoretical background}
\label{sec:theory}
In this section, we introduce the theoretical formalism and key quantities for weak lensing and the intrinsic alignment of galaxies. 

\subsection{Weak gravitational lensing}
\label{subsec:wl}
We assume a Friedmann-Lemaître-Robertson-Walker universe, characterized by weak scalar inhomogeneous perturbations expressed in terms of the Newtonian gravitational potential $\Phi$. In the absence of anisotropic stress, the two Bardeen potentials are equal, and the metric is given by:
\begin{equation}
\label{eq:flrw}
{\rm d}s^2 = -\left(1 + \frac{2\Phi}{c^2}\right) \, c^2{\rm d}t^2 + a^2(t) \left(1 - \frac{2\Phi}{c^2}\right) \, \delta_{ij} \, {\rm d}x^i {\rm d}x^j \, ,
\end{equation}
where $\delta_{ij}$ is the Kronecker delta, $c$ is the speed of light, and $a(t)$ is the scale factor. Given a ray of light that travels from redshift $z_{\rm s}$ to the observer, its trajectory is deflected by the gravity of the intervening matter field. This deflection is expressed through the lens equation: 
\begin{equation}
\label{eq:lens_compact}
\bm{\beta} = \bm{\theta} - \bm{\alpha} \, ,
\end{equation}
which relates the true (source) position $\bm{\beta}$, the observed (image) position $\bm{\theta}$, and the deflection angle $\bm{\alpha}$. In our context, $\bm{\alpha}$ can be written explicitly as follows, giving the gravitational lens equation:
\begin{equation}
\label{eq:lens}
\bm{\beta}(\bm{\theta}, z_{\rm s})
 = \bm{\theta} - \frac{2}{{\rm c}^2} \int^{\chi_{\rm s}}_0 {\rm d} \chi_{\rm d} \frac{f_{\rm ds}}{f_{\rm d} f_{\rm s}} \nabla_{\bm{\beta}} \Phi (\bm{\beta}(\bm{\theta}, \chi_{\rm d}), \chi_{\rm d}, z_{\rm d}) \, .
\end{equation}
Here, we have defined the gradient with respect to the angular position on the sky $ \nabla_{\bm{\beta}} $, the comoving line-of-sight distance $ \chi $, and the comoving angular diameter distance $ f_{K}(\chi) $. The subscripts ``s'' and ``d'' denote the source and the deflector (i.e.~the lens), respectively. Consequently, the geometric factors are given by $ f_{\rm ds} = f_{K}(\chi_{\rm s} - \chi_{\rm d}) $, $ f_{\rm d} = f_{K}(\chi_{\rm d}) $, and $ f_{\rm s} = f_{K}(\chi_{\rm s}) $.

To characterise the WL signal at a given point in space, we assume that source galaxies have a small angular extent on the sky, and are subject to small lensing induced deflections. This allows us to write down the Jacobian matrix of the lens equation and perform a linear expansion. This matrix can be expressed in the following form, and is referred to as the distortion matrix:
\begin{equation}
\label{eq:jacobian}
\frac{\partial \bm{\beta}}{\partial \bm{\theta}}
 = 
 \begin{pmatrix}
1 - \kappa - \gamma_1 & -\gamma_2 - \omega \\
-\gamma_2 + \omega    & 1 - \kappa + \gamma_1
\end{pmatrix}
\, .
\end{equation}
Here the convergence, $\kappa$, is a scalar that measures the isotropic lensing distortion, which corresponds to a uniform scaling of the image. The rotation, $\omega$, is a scalar that expresses the rigid rotation of the image about its center; it can be shown that in the context of WL one can typically assume $\omega \approx 0$. The shear, $\gamma = \gamma_1 + {\rm i}\gamma_2$, is a spin-2 quantity that describes the anisotropic deformation of the source, which corresponds to elongation in a specific direction, as dictated by the individual components $\gamma_1$ and $\gamma_2$. In this work we adopt the convention that $\gamma_1$ corresponds to east-west elongation and $\gamma_2$ corresponds to northeast-southwest elongation. From here on, we will refer to the gravitationally induced (extrinsic) shear as $\gamma_{\rm G}$, and distinguish it from the intrinsic shear $\gamma_{\rm I}$, which we introduce in the following paragraph.

\subsection{Intrinsic ellipticity of galaxies}
\label{subsec:ellipticity}
The intrinsic three-dimensional shape of galaxies can be approximated as a triaxial ellipsoid. The projection of such an object onto the celestial sphere (described in detail in Section~\ref{subsec:intr_shear}) results in an ellipse with major axis $a$, minor axis $b$, and orientation angle $\phi$, defined as  East of North. This lets us introduce another spin-2 field, the ellipticity\footnote{Note that there is more than one possible definition of the ellipticity, for a discussion we refer the interested reader to section 2.2 of \citet{Lamman2024}.}:
\begin{equation}
\label{eq:ellipticity}
\epsilon \equiv \frac{a^2 - b^2}{a^2 + b^2} \exp{(2{\rm i}\phi)}
\, ,
\end{equation}
which can be converted to the intrinsic shear through:
\begin{equation}
\label{eq:response_factor}
\gamma_{\rm I} = \frac{\epsilon}{2 {\cal R}}
\, .
\end{equation}
Here the responsivity factor ${\cal R}$ accounts for the average response of the ellipticity to a given shear. Consistently with D23, in the following we assume ${\cal R} = 1$, justified by the assumption that galaxy isophotes are elliptical and the absence of measurement noise \citep[see e.g.][]{Bernstein2002}. Finally, under the assumption that the gravitational shear is sufficiently small, one can obtain the total (observed) shear $\gamma_{\rm O}$ simply by adding the intrinsic and the (extrinsic) gravitational components:
\begin{equation}
\label{eq:shear_tot}
\gamma_{\rm O} \approx \gamma_{\rm I} + \gamma_{\rm G}
\, ,
\end{equation}
which is the relation employed in our work.

\subsection{Correlation function and power spectrum}
\label{subsec:ps_corrfunc}

The two-point shear correlation function is the most well studied and widely used WL summary statistic. For two generic shear fileds $A$ and $B$, it is defined as:
\begin{equation}
\label{eq:xi_pm}
\xi_{\pm, AB}(\theta) = \langle \gamma_{A, +}(\bm{\theta}') \gamma_{B, +}(\bm{\theta}'+\bm{\theta}) \rangle 
\pm \langle \gamma_{A, \times}(\bm{\theta}') \gamma_{B, \times}(\bm{\theta}'+\bm{\theta}) \rangle \, ,
\end{equation}
where $\gamma_+$ and $\gamma_\times$ are the tangential and cross components of the shear with respect to the direction connecting a pair of galaxies separated by an angular distance\footnote{In the context of the correlation function, $\theta$ denotes a scalar separation angle between two objects on the celestial sphere. This should be distinguished form $\bm \theta$, which refers to a two-dimensional position vector on the sky, whose components are $\theta_1$ and $\theta_2$.} $\theta$. Given the Cartesian shear components $\gamma_1$ and $\gamma_2$, these projections are defined as:
\begin{subequations}
\begin{align}
\label{eq:gamma_proj}
\gamma_+ &= -\left( \gamma_1 \cos 2\varphi + \gamma_2 \sin 2\varphi \right) \, , \\
\gamma_\times &= -\left( -\gamma_1 \sin 2\varphi + \gamma_2 \cos 2\varphi \right) \, ,
\end{align}
\end{subequations}
where $\varphi$ is the polar angle of the separation vector between the two galaxies (for a schematic visualization, see e.g. Figure 2 of D23).

The correlation function is particularly useful for comparing observational data with theoretical models in real space, as it can be directly estimated from galaxy shape measurements. In this work, we compute its $\mathrm{GG}$ (gravitational), $\mathrm{II}$ (intrinsic), and $\mathrm{GI}$ (cross) components.

The Fourier counterpart of the correlation function is the angular power spectrum $ C_{AB}(\ell) $, the two are related via the Hankel transforms:
\begin{equation}
\label{eq:hankel_transform}
\xi_{+/-, AB}(\theta) = \int_0^{\infty} \frac{\ell \, {\rm d}\ell}{2\pi} \, C_{AB}(\ell) \, J_{0/4}(\ell \theta) \, ,
\end{equation}
where $J_0$ and $J_4$ are the zeroth- and fourth-order Bessel functions, respectively \citep[see e.g.][]{schneider2002}. 

In this study we will focus on the shear correlation function and on the convergence power spectrum. It can be shown that under the Limber~\citep[][]{limber1953, loverde2008}, flat-sky, and Born~\citep[see e.g.][]{Ferlito2024} approximations, the angular power spectra of the shear (E and B modes), convergence, and rotation, are connected as follows:
\begin{subequations}
\label{eq:shearcl}
\begin{align}
  & C_{\gamma}^{\rm (EE)}(\ell) = C_{\kappa}(\ell) \, , \\
  & C_{\gamma}^{\rm (BB)}(\ell) = C_{\omega}(\ell) = 0 \, .  
\end{align}
\end{subequations}
This justifies the interchangeability of the convergence and shear power spectrum under the assumption of negligible shear B-modes~\citep[see e.g.][]{Kilbinger2015}.

\medskip
\noindent After  recapitulating  some fundamentals of WL theory, we now present our methods in the next section.


\section{Methods}
\label{sec:methods}

\begin{figure*}
    \centering
    \includegraphics[width=\textwidth]{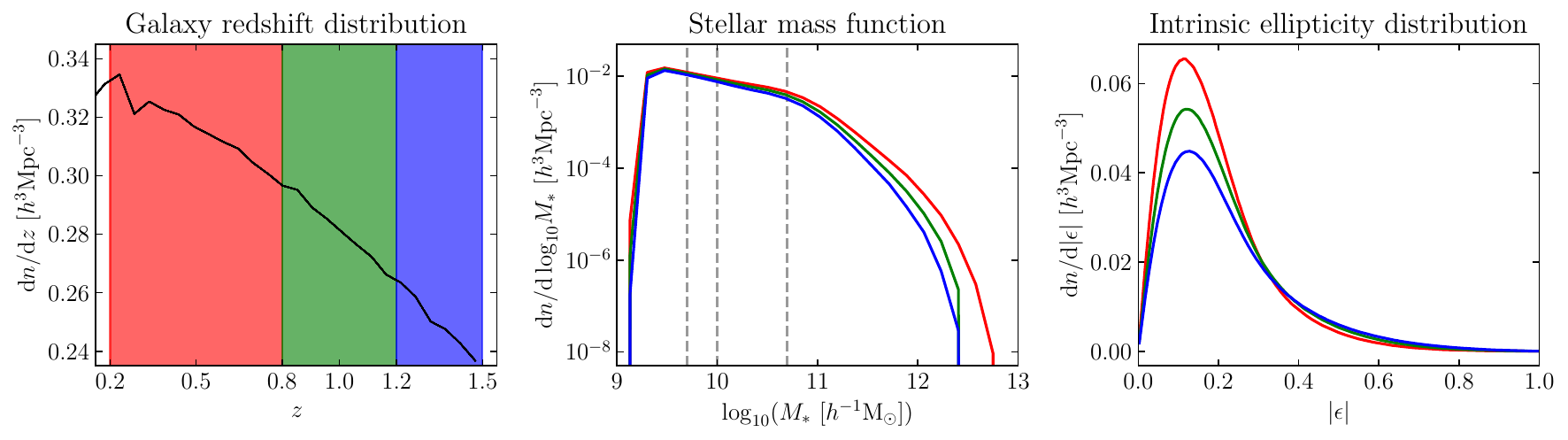}
    \caption{The left, center and right panels show the redshift distribution, the mass function and the absolute ellipticity distribution of galaxies in our catalogue, respectively. The shaded regions in the left panel refer to the three redshift bins used in this work and are consistent with the colours adopted in the two rightmost panels. The grey dashed lines in the central panel refer to the minimum stellar mass cuts of $[5\times10^9, 1\times10^{10}, 5\times10^{10}] \, \Msh$ applied to the catalogue in Section~\ref{subsec:mass_dependence}. We observe that, as expected, the galaxy number density decreases with increasing redshift. We also note that, as the redshift decreases, the mass function grows consistently with hierarchical structure formation, and that the galaxy ellipticities tend to move toward lower values (i.e. rounder shapes).}
    \label{fig:redshift_mass_ellipticity_dist}
\end{figure*}

\subsection{Simulations}
\label{subsec:mtng}
In our work, we use the flagship, full-hydrodynamics run of the MillenniumTNG (MTNG) project, a state-of-the-art suite that includes both large dark matter-only simulations (some of which include massive neutrinos) and corresponding fully hydrodynamical simulations in comparatively large volumes. The simulation we employ, hereafter MTNG740, was performed using the {\sc Arepo} code \citep{Springel2010, Weinberger2020} and contains $4320^3$ dark-matter particles with mass $1.7\times10^8\Msol$, and $4320^3$ gas cells, each with an initial mass of $3.1\times10^7\Msol$ in a periodic box with comoving side-length $L_{\mathrm{box}}=500 \Mpch \approx 740\ \mathrm{Mpc}$. The MTNG740 simulation adopts the same baryonic physics modelling as the {\sc IllustrisTNG} project \citep{Nelson2018, Springel2018, Marinacci2018, Pillepich2018, Naiman2018, Pillepich2019, Nelson2019a, Nelson2019b}, modulo some minor changes\footnote{Magnetic fields were not included and the metallicity tracking was simplified in order to reduce the memory consumption such that the production run could fit into the available memory.}. 

The cosmological parameters of MTNG740 were set to the \cite{Planck2015} cosmology, the same as adopted in {\sc IllustrisTNG}: $\Omega_{\rm m} = \Omega_{\rm cdm} + \Omega_{\rm b} = 0.3089$, $\Omega_{\rm b} = 0.0486$, $\Omega_\Lambda = 0.6911$, $h = 0.6774$, $\sigma_8 = 0.8159$ and $n_s=0.9667$. The initial conditions were set at $z=63$ and were generated via second-order Lagrangian perturbation theory with an updated version of the {\small N-GENIC} code as embedded in the {\small GADGET-4} \citep{Springel2021} code. We direct the reader to \citet{Pakmor2023} for further details regarding the full-hydrodynamic run, and to \citet{Aguayo2023} for an overview of the simulation suite. 


\subsection{Generation of the galaxy catalogue}
\label{subsec:gal_cat_generation}

Our galaxy catalogue is generated by running an updated version of the Friends-of-Friends (FoF) group finding algorithm \citep{Davis1985} combined with the {\sc Subfind-HBT} substructure finder codes present in {\small GADGET-4}. These routines have been specifically optimised for efficiency on lightcones, where the output structure consists of a succession of consecutive redshift slices. In this work, we decided to use the particle lightcone with the combination of sky coverage and redshift depth that best matches the specifications of a stage IV survey. The chosen lightcone corresponds to one octant of the sky (coordinates $x > 0$, $y > 0$, $z > 0$) for redshifts $0 \leq z \leq 1.5$, reaching comoving distance $\approx 3050\Mpch$ \citep[see Section 3.5 of][for a comprehensive list of the available MTNG lightcones]{Aguayo2023}. For this lightcone, the raw particle data amounts to over 540 TB, requiring significant computational resources to process. From this, by means of the FoF algorithm, we identified more than $1.30 \times 10^{10}$ halos, and, by means of the {\sc Subfind-HBT} algorithm, we identified more than $1.26 \times 10^{10}$ substructures.

\subsection{Galaxy selection and redshift distribution}
\label{subsec:gal_cat_selection}

In order to cover a galaxy stellar mass range that is as close as possible to that which is observed with stage IV surveys, we select our galaxies as the subhalos with at least 100 stellar particles and a minimum total mass of $1.0\times10^{10}\Msol$. This results in a galaxy population with a minimum stellar mass of $\approx 1.0\times10^9\Msol$, as one can see in Figure~\ref{fig:redshift_mass_ellipticity_dist}, where we show the stellar mass function of our catalogue in the three redshift bins described in the following paragraph. We point out that the same stellar particle threshold has been applied in \citet{Hilbert2017}, and we refer the reader to \citet{Chisari2015} and \citet{Velliscig2015a} for details on measurement errors of ellipticity distributions as a function of particle number.


With the selection described above, the resulting galaxy redshift distribution is shown in the left panel of Figure~\ref{fig:redshift_mass_ellipticity_dist}. Our catalogue features a total of $\approx2.10\times10^8$ galaxies, corresponding to an angular source density of $\approx11.31$ galaxies per square arcminute. In order to investigate the redshift dependence of the impact of IA on WL observables, we partition our catalogue into three tomographic redshift bins with edges at $z = [0.2, 0.8, 1.2, 1.5]$, highlighted by the shaded regions in the left panel of Figure~\ref{fig:redshift_mass_ellipticity_dist}. The corresponding comoving distance width for each of the bins is $\Delta\chi \approx [1371, 676, 416] \Mpch$. This choice results in each bin containing a similar number of galaxies, and thus gives a constant noise level between different redshift bins.

\subsection{Computation of intrinsic ellipticity}
\label{subsec:intr_shear}

In numerical simulations, galaxies and haloes are typically represented by an ensemble of $N$ particles. A common way to quantitatively describe the three-dimensional shape of these objects is to consider the inertia tensor $I$, which sums over all particle distances from the center associated with a given object and is given by the expression:
\begin{equation}
\label{eq:inertia_tensor}
I_{ij}=\frac{1}{W}\sum_{k=1}^N w^k x_i^k x_j^k
\, ,
\end{equation}
where $x$ represents the distance of a given particle from the centre of the object, with the indices $i$ and $j$ running over the three spatial components, $i,j \in (x,y,z)$, and $w_k$ are weight factors (e.g.~mass or luminosity of the particles) whose sum is $W$. We now want to compute the 2D ellipticity of the galaxy as seen on the celestial sphere, defined in Equation~(\ref{eq:ellipticity}). To do so, we need to project $I$ onto a two-dimensional surface perpendicular to the line of sight $\hat{\bm n}$ at the galaxy angular position ${\bm \theta}$. Similarly to \citet{Tsaprazi2022}, we define an orthonormal basis $(\hat{\bm n}, \hat{{\bm\theta}_1}, \hat{{\bm\theta}_2})$, where  
\begin{subequations}
  \begin{align}
    \hat{\bm{\theta}}_1 &= \cos{\theta_1} \cos{\theta_2} \hat{\bm x} + \cos{\theta_1} \sin{\theta_2} \hat{\bm y} - \sin{\theta_1} \hat{\bm z} \, , \\
    \hat{\bm{\theta}}_2 &= - \sin{\theta_2} \hat{\bm x} + \cos{\theta_2} \hat{\bm y} \, ,
  \end{align}
\end{subequations}
with $(\theta_1$, $\theta_2)$ being the colatitude and longitude of the galaxy, and $(\hat{\bm x}, \hat{\bm y}, \hat{\bm z})$ is the orthonormal basis of the Cartesian coordinates. 

Following \citet{Schmidt2012} and \citet{Schmidt2015} we then define the linear combinations:
\begin{equation}
\label{eq:lin_comb}
{\bm m}_{\pm} \equiv({\hat{\bm{\theta}}_2} \mp {\rm i}{\hat{\bm{\theta}}_1}) / \sqrt{2}
\, ,
\end{equation}
where ${\rm i}$ is the imaginary unit. This allows us to decompose the inertia tensor into the spin-2 field $I_{\pm} = I_1 \pm {\rm i} I_2$ according to
\begin{equation}
\label{eq:inertia_tensor_rotated_basis}
I_{\pm} = \sum_{i=1}^3 \sum_{j=1}^3 m^i_{\mp} m^j_{\mp} I_{ij}
\, .
\end{equation}
In this way, after having also defined the normalization factor
\begin{equation}
\label{eq:inertia_tensor_norm}
I_{\rm norm} = \sum_{i=1}^3 \sum_{j=1}^3 m^i_{-} m^j_{+} I_{ij}
\, ,
\end{equation}
we can write the ellipticity projected onto the sky as:
\begin{equation}
\label{eq:projected_ellipticity}
\epsilon = \epsilon_1 \, {\pm} \, {\rm i} \epsilon_2  = \frac{I_1 \pm {\rm i} I_2}{I_{\rm norm}}
\, .
\end{equation}

In this work, consistent with D23, we will use the V-band luminosity as a weight for the inertia tensor, which leads to the following specialisation of equation~(\ref{eq:inertia_tensor}):
\begin{equation}
    q_{ij}^{(V)}=\frac{1}{L^{(V)}}\sum_k^N L^{k, (V)} \left(x^k-\overline{x}^{(V)}\right)_i \left(x^k-\overline{x}^{(V)}\right)_j
    \label{eq:luminosity_tensor} \, ,
\end{equation}
where $L^{k, (V)}$ and $L^{(V)}$ are the V-band luminosity of the $k$-th stellar particle and of the whole galaxy, respectively; and ${x}^k$ and $\overline{x}^{(V)}$ are the position vector of the $k$-th stellar particle and the V-band luminosity weighted centre of the galaxy, respectively.

As a sanity check, in the right panel of Figure~\ref{fig:redshift_mass_ellipticity_dist} we show the ellipticity modulus distribution in the three redshift bins we defined in Sec. \ref{subsec:gal_cat_selection}. We find that both the shape distribution and the redshift evolution of the distribution are qualitatively consistent with similar studies \citep{Samuroff2021, Chisari2015, zhang2022, Lee2025}, with an increasing fraction of rounder object as the redshift is decreasing. This trend can be mainly interpreted as a consequence of phenomena like dynamical relaxation \citep[see e.g.][]{lynden1967, binney2008} and mergers \citep[see e.g.][]{negroponte1983, naab2003}, both of which tend to isotropize galaxy shapes over time. We note that the observed trend toward rounder galaxy shapes at lower redshift, while qualitatively consistent with expectations from the aforementioned effects, may be affected by numerical limitations. In particular, the relatively low particle count per galaxy in MTNG740 can lead to artificially short relaxation times, potentially exaggerating this effect. A natural extension would be to perform higher-resolution zoom-in simulations of selected galaxies within the MTNG volume to assess how shape distributions evolve with improved resolution.

\subsection{Computation of galaxy image positions and gravitational shear}
\label{subsec:wl_shear}

To compute the gravitational shear and the image (observed) positions of the galaxies, we start by producing a collection of full-sky shear and deflection HEALPix maps \citep{Gorsky2005} using the {\small DORIAN} ray-tracing code \citep{Ferlito2024}. We compute these maps for slices with a thickness of $100 \Mpch$, starting at a comoving distance of $100 \Mpch$  from the observer ($z\approx0.03$), up to $3.1\, h^{-1}\mathrm{Gpc}$ ($z\approx1.55$). We use a HEALPix tessellation with $N_{\rm side}=8192$, which yields an angular resolution of 0.43 arcmin.

A deflection map relates the image position of each ray, located at the pixel centre of a regular HEALPix grid, to the respective position on the source plane, where we note that source positions generally do not follow a regular arrangement. We can use the deflection map to compute the image position of galaxies in our catalogue, starting from their source (true) positions. For a given galaxy source position ${\bm \beta}_{\rm g}$, one can define a triangle whose vertices are neighbouring rays\footnote{Note that the rays have to be neighbours in the image plane.} and that contains ${\bm \beta}_{\rm g}$; once such a triplet of rays is found, one can perform a barycentric interpolation \citep[see e.g. section 21.3 of][]{Press2007} from the source plane to the image plane to estimate the observed galaxy position ${\bm \theta}_{\rm g}$ \citep[see e.g. Figure~7 of][for a visualization of this interpolation scheme]{Hilbert2009}. Operating on the curved sky, the algorithm we implemented for the galaxy image search is similar in many aspects to the one of \citet[][we refer the interested reader to section 3.3 of this work for further details]{Becker2013}. One difference is that our code performs a KD-tree-based neighbour search, which looks for the 30 closest rays on the source plane for each galaxy\footnote{We found that looking for the $\approx10$--$20$ closest rays already gives convergent results.}. 

For each galaxy, we first compute the image position using the two deflection maps corresponding to the neighbouring source planes that enclose the galaxy's redshift; then we perform a linear interpolation between these two to obtain our final estimate of the image position. We note that galaxies for which multiple image positions are found are excluded from the catalogue, as they are likely subject to strong lensing, and therefore would be excluded in a WL lensing survey.

Once the observed galaxy positions are determined, we apply non-uniform fast Fourier transform (NUFFT)  interpolation \citep{Fessler2003, Barnett2019, Reinecke2023, Ferlito2024} on HEALPix shear maps to compute the gravitational shear at those positions. Again, we evaluate this quantity at the two maps with source redshifts that enclose each galaxy and linearly interpolate the shear values to obtain the final estimate. Compared to the more conventional approaches like nearest-grid-point or bilinear interpolation on the HEALPix grid, NUFFT offers both higher accuracy and efficiency when evaluating the shear at arbitrary positions.

\subsection{From shear to convergence maps}
\label{subsec:gamma_to_kappa}

\begin{figure*}
\begin{adjustbox}{max width=\textwidth}
\begin{tikzpicture}
\node[anchor=south west, inner sep=0] (A) at (0,0) {\includegraphics[height=0.9\textheight,page=1]{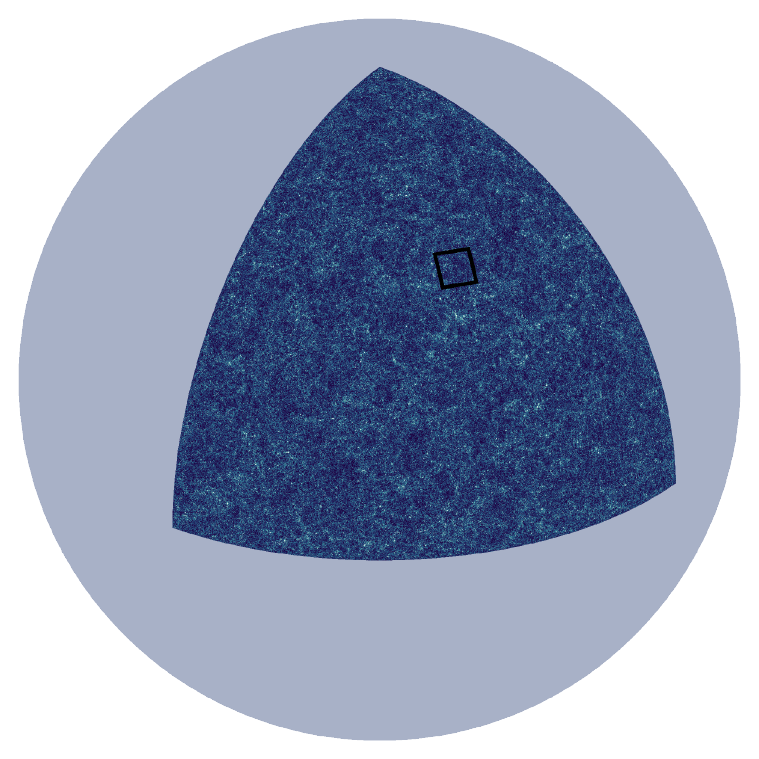}};

\node[anchor=south west, inner sep=0] (B) at ([xshift=0.5cm]A.south east) {\includegraphics[height=0.9\textheight,page=1]{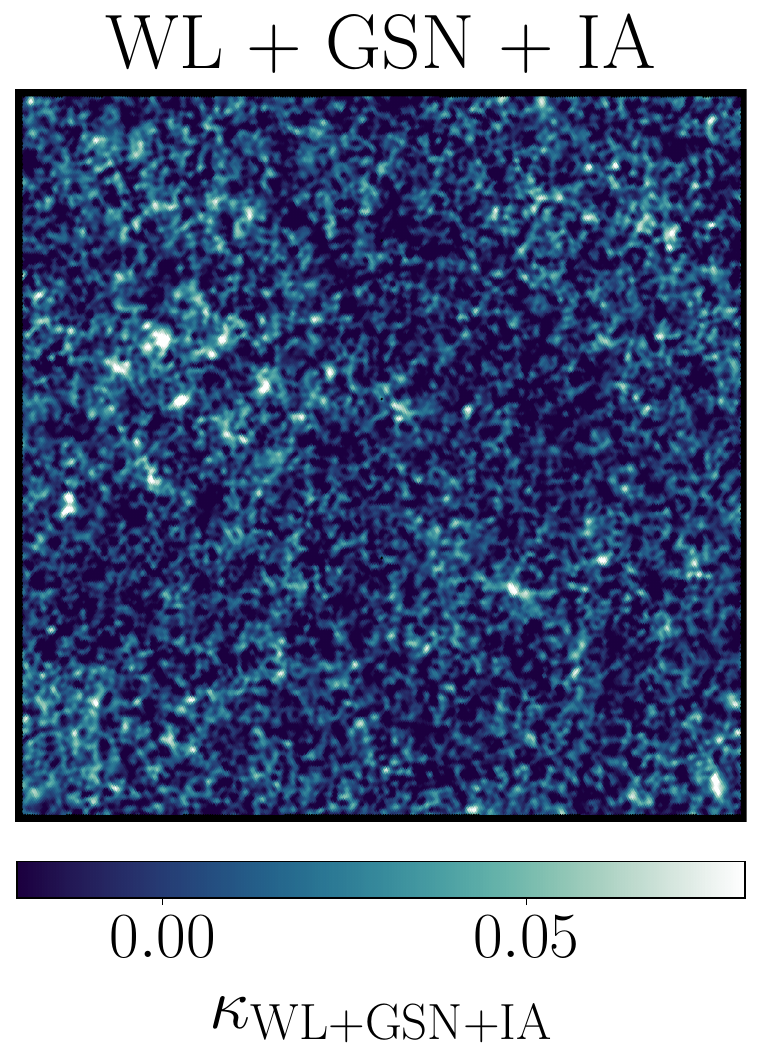}};
\node[anchor=south west, inner sep=0] (C) at ([xshift=0.5cm]B.south east) {\includegraphics[height=0.9\textheight,page=1]{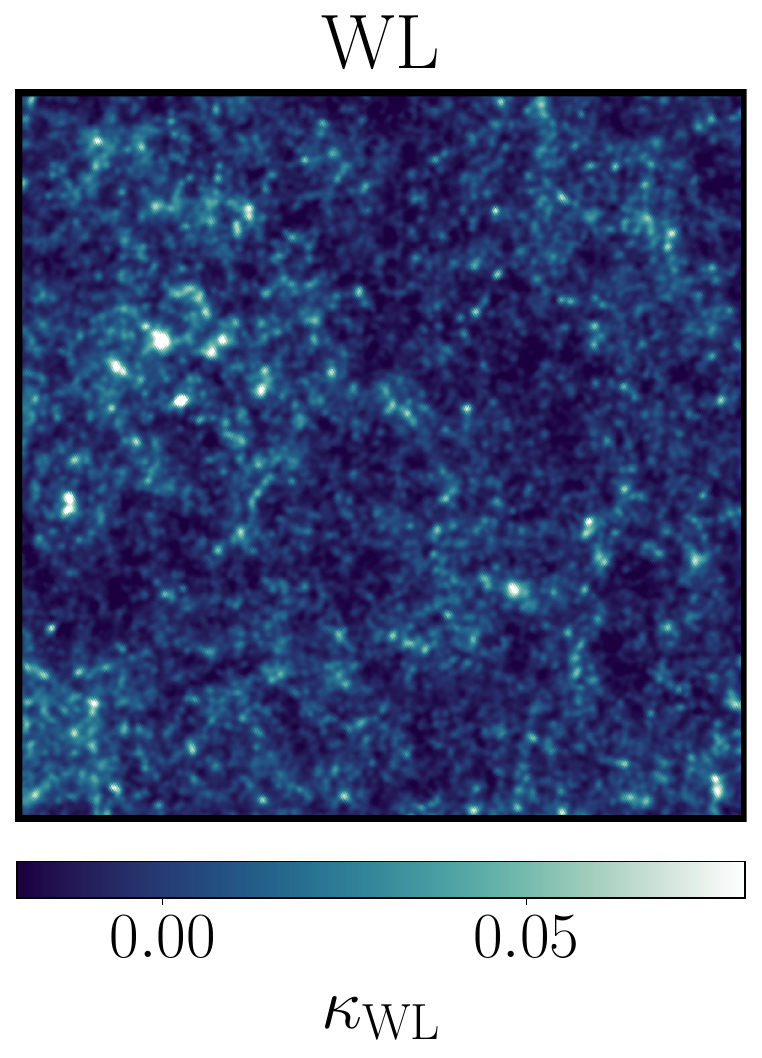}};
\node[anchor=south west, inner sep=0] (D) at ([xshift=0.5cm]C.south east) {\includegraphics[height=0.9\textheight,page=1]{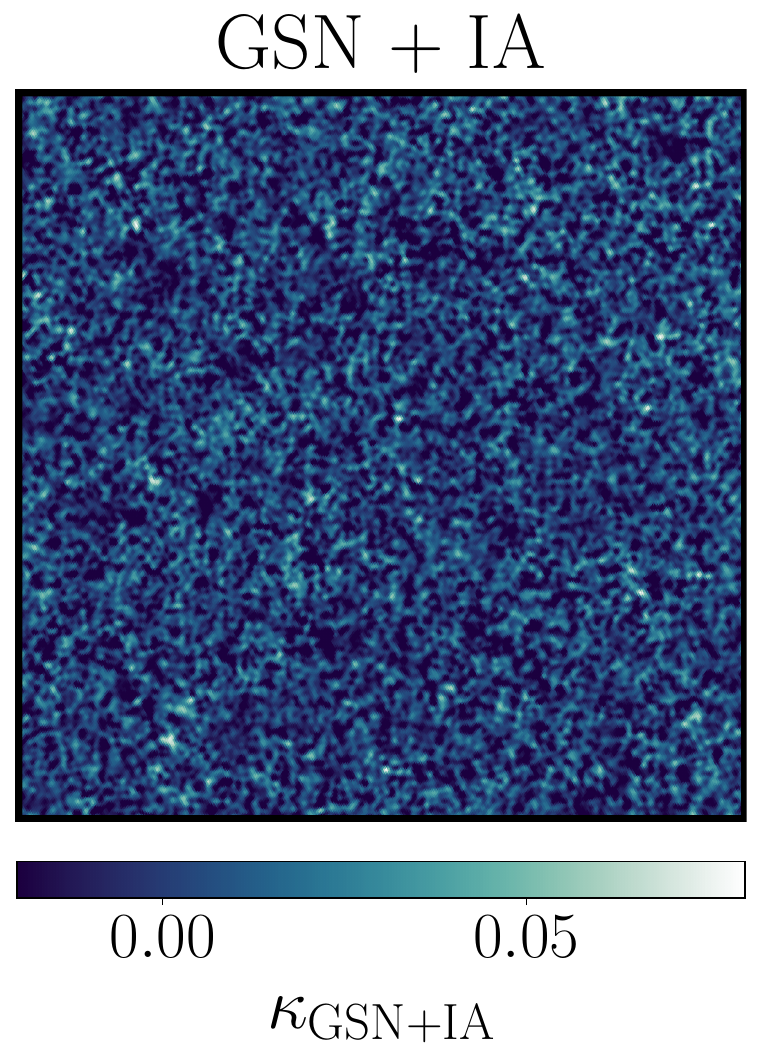}};
\node[anchor=south west, inner sep=0] (E) at ([xshift=0.5cm]D.south east) {\includegraphics[height=0.9\textheight,page=1]{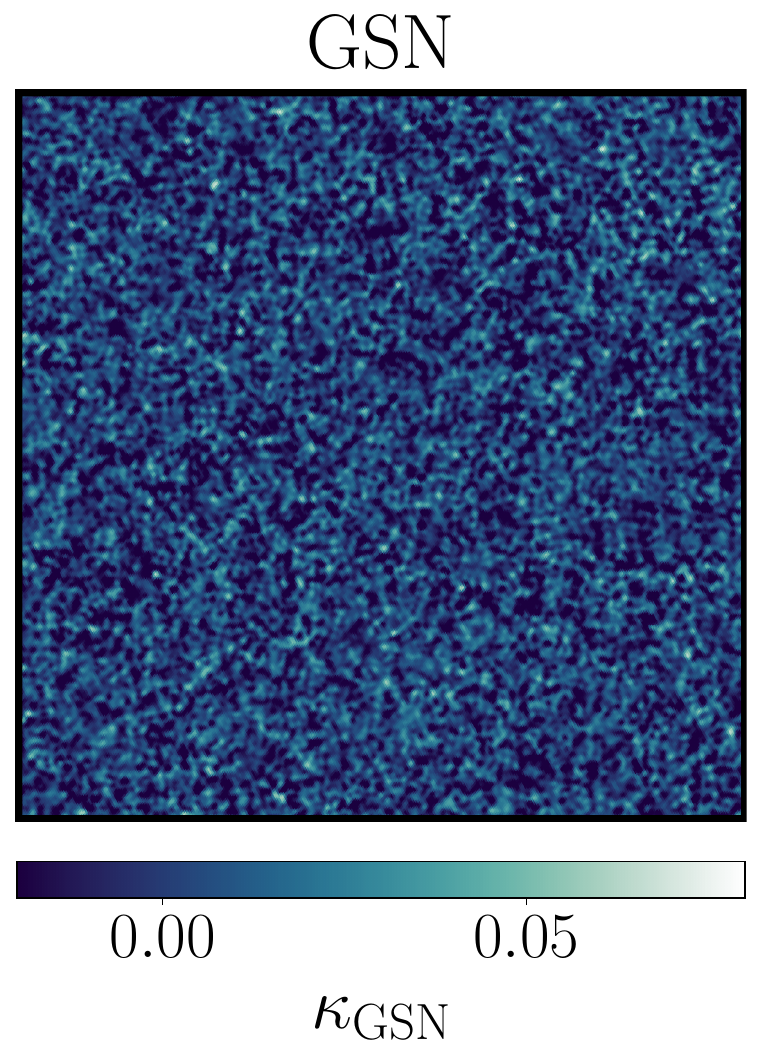}};

\draw[black, line width=3.5pt, -] ([xshift=-9.268cm,yshift=14.339cm]A.south east) -- ([xshift=0.34cm,yshift=19.698cm]B.south west);
\draw[black, line width=3.5pt, -] ([xshift=-9.007cm,yshift=13.397cm]A.south east) -- ([xshift=0.349cm,yshift=4.881cm]B.south west);
\end{tikzpicture}
\end{adjustbox}
\caption{{\it From left to right:} full ``WL + GSN + IA'' (Weak Lensing + Galaxy Shape Noise + Intrinsic Alignment) map displayed as an octant of the sky with a zoom into a $10 \times 10 \, \mathrm{deg}^2$ square patch, followed by zoomed patches of the same area, for the ``WL'', ``GSN + IA'', and ``GSN'' cases. The maps are computed considering source galaxies in our central redshift bin, $0.8<z<1.2$. For illustrative purposes we have smoothed the maps using a Gaussian symmetric beam characterised by a standard deviation of $1 \, \mathrm{arcmin}$. We see that the IA contribution in the ``GSN + IA'' panel leads to weak correlations with the WL signal, while the ``GSN'' panel appears consistent with white noise.}
\label{fig:octant_zoom}
\end{figure*}

With the methodology outlined above, we have built a catalogue containing, amongst other quantities, each galaxy's observed angular position, intrinsic shear, and gravitational shear. Our aim is now to compute convergence maps on a regular (HEALPix) grid starting from the value of the shear at the observed galaxy position. To perform this transformation we use the relationship between $\kappa$ and $\gamma$ in Fourier space \citep{Kaiser1993}, known as the Kaiser-Squires transformation. In particular, we first extract the spherical harmonics coefficients $\gamma_{\ell m}$ from the shear map with the \texttt{pseudo\_analysis\_general} routine\footnote{This routine can be seen as a generalized version of the HEALPix \texttt{alm2map} routine, which accepts as input a generic, non-uniform, distribution of points on the sphere.}, which is part of the {\sc ducc} library \citep{Reinecke2020}, and then use the relation
\begin{equation}
\label{eq:kappa_lm}
\kappa_{\ell m} = - \gamma_{\ell m} / \sqrt{(\ell+2) (\ell-1) / (\ell (\ell+1))}  \, 
\end{equation}
to obtain the spherical harmonics coefficients of the convergence \citep[we refer the interested reader to][for a derivation of this relation]{Hu2000}. We then use the HEALPix \texttt{alm2map} routine to compute the convergence map on the regular HEALPix grid.

To directly test the impact of IA on higher-order statistics measured in the $\kappa$ field, we compute three main types of convergence maps, depending on the shear field we use as input for the above equation. The first one, labelled ``WL + GSN + IA'', is obtained by using the total shear given by equation~(\ref{eq:shear_tot}); this field contains the WL signal as well as the galaxies' intrinsic shear contribution, introduced in equation~(\ref{eq:response_factor}). By plugging equation~(\ref{eq:ellipticity}) into equation~(\ref{eq:response_factor}), one can see that the intrinsic shear, represented by a complex quantity, can be split into its modulus (i.e. shape), which gives rise to the \textit{galaxy shape noise} component \citep[GSN, see e.g.][]{Kaiser1995, Bernstein2002}, and orientation, which contains information about the intrinsic alignment. 

The second type of convergence map, labeled ``WL + GSN'', is obtained by adding the gravitational shear and a modified version of the intrinsic shear field, where for each galaxy the shape is preserved, but the orientation has been rotated by a random angle. This preserves the shape noise contribution, but erases the intrinsic alignment. 

Finally, the third type is obtained by considering only the gravitational shear component, and is referred to as ``WL''. Additionally, one can compute the ``GSN + IA'' and ``GSN'' cases, by using the original intrinsic shear and the intrinsic shear with randomized orientations as inputs respectively. 

We note that given that the WL signal is only evaluated at (observed) galaxy positions, the convergence maps generated here also include contamination of the lensing signal from source clustering. 

Figure~\ref{fig:octant_zoom} shows a visualisation of the different types of convergence maps described above, for the central redshift bin $0.8<z<1.2$. For illustrative purposes we have smoothed the maps using a Gaussian symmetric beam characterised by a standard deviation of $1 \, \mathrm{arcmin}$. The first panel shows the full ``WL + GSN + IA'' map displayed as an octant of the sky. Subsequent panels show a zoom into a $10 \times 10 \, \mathrm{deg}^2$ square patch, starting with a zoom of the same ``WL + GSN + IA'' map. We then show zoomed patches of the same area, for the ``WL'', ``GSN + IA'', and ``GSN'' cases. The first zoomed panel (``WL + GSN + IA'') clearly depicts a noisy realisation of large-scale-structure features, which can be visually verified by comparing it to the adjacent panel of ``WL'', where only the true WL signal is displayed. The next panel shows the ``GSN + IA'' map, where weak correlations with the `WL'' map can be observed. This can be attributed to the IA signal in the ``GSN + IA'' map, since the IA signal is a direct consequence of the large-scale structure. Finally, the last panel shows the ``GSN'' map, which is visually consistent with a white noise field.

\subsection{Computation of WL and IA statistics}
\label{subsec:statistics}

Here we outline the numerical procedures and binning specifications used when measuring each of the WL statistics studied in this work. The shear correlation function is computed with the {\sc TreeCorr} code \citep{Jarvis2015}, and binned into $40$ equally spaced logarithmic bins in the range $\theta \in [1, 400]$ arcmin. The convergence power spectrum is computed using the HEALPix routine \texttt{anafast}, and binned into 50 equally spaced logarithmic bins in the range $\ell \in [1,1.2 \times 10^4]$. Before computing probability distribution functions (PDFs) for convergence, peaks and minima, every convergence map is smoothed with a Gaussian symmetric beam characterised by a standard deviation of $2 \, \mathrm{arcmin}$ using the HEALPix \texttt{smoothing} routine. For the convergence PDF, we bin the pixels into 50 linearly spaced bins in the range $\kappa \in [-0.07, 0.12]$. Peaks and minima are identified as the pixels that are greater or smaller than their 8 neighbours, respectively\footnote{In the HEALPix tessellation, every pixel has 8 neighbours, except for a small minority of pixels, for which it can be 7 or 6.}, which are retrieved using the HEALPix \texttt{get\_all\_neighbours} routine. For the peaks, we use 30 linearly spaced bins in the range $\kappa \in [-0.1, 0.2]$, while for the minima we use 25 linearly spaced bins in the range $\kappa \in [-0.07, 0.05]$.


\section{Results}
\label{sec:results}

\subsection{Comparison to theory}
\label{subsec:comp_theory}

\begin{figure*}
    \centering
    \includegraphics[width=\textwidth]{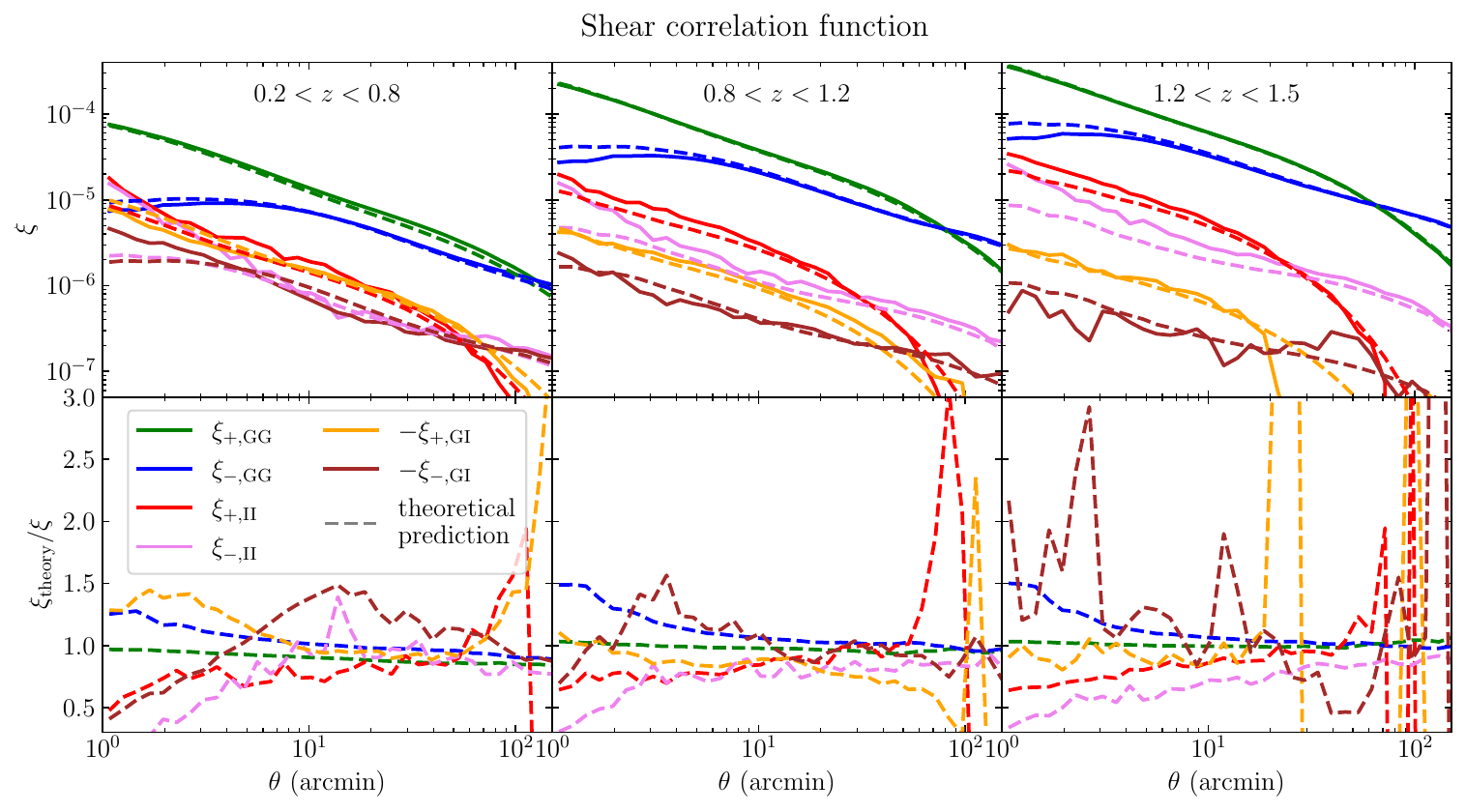}
    \caption{Comparison between the shear correlation function as computed on our results and its theoretical prediction, obtained by means of the {\sc CCL} library \citep{Chisari2019}. Each column refers to a different redshift bin. The top panels show the shear correlation function, while the bottom panels give the ratio of the theoretical prediction over our results. In the case of  $\xi_{\mathrm{II}}$ and $\xi_{\mathrm{GI}}$ we use the Nonlinear Alignment model (NLA) for the theory prediction, which provides good overall agreement with the measured signals, with best-fit alignment amplitudes around $A_1 \approx [2.06, 2.55, 3.52]$ for bins with increasing redshift.}
    \label{fig:corr_func_check}
\end{figure*}

\begin{figure*}
    \centering
    \includegraphics[width=\textwidth]{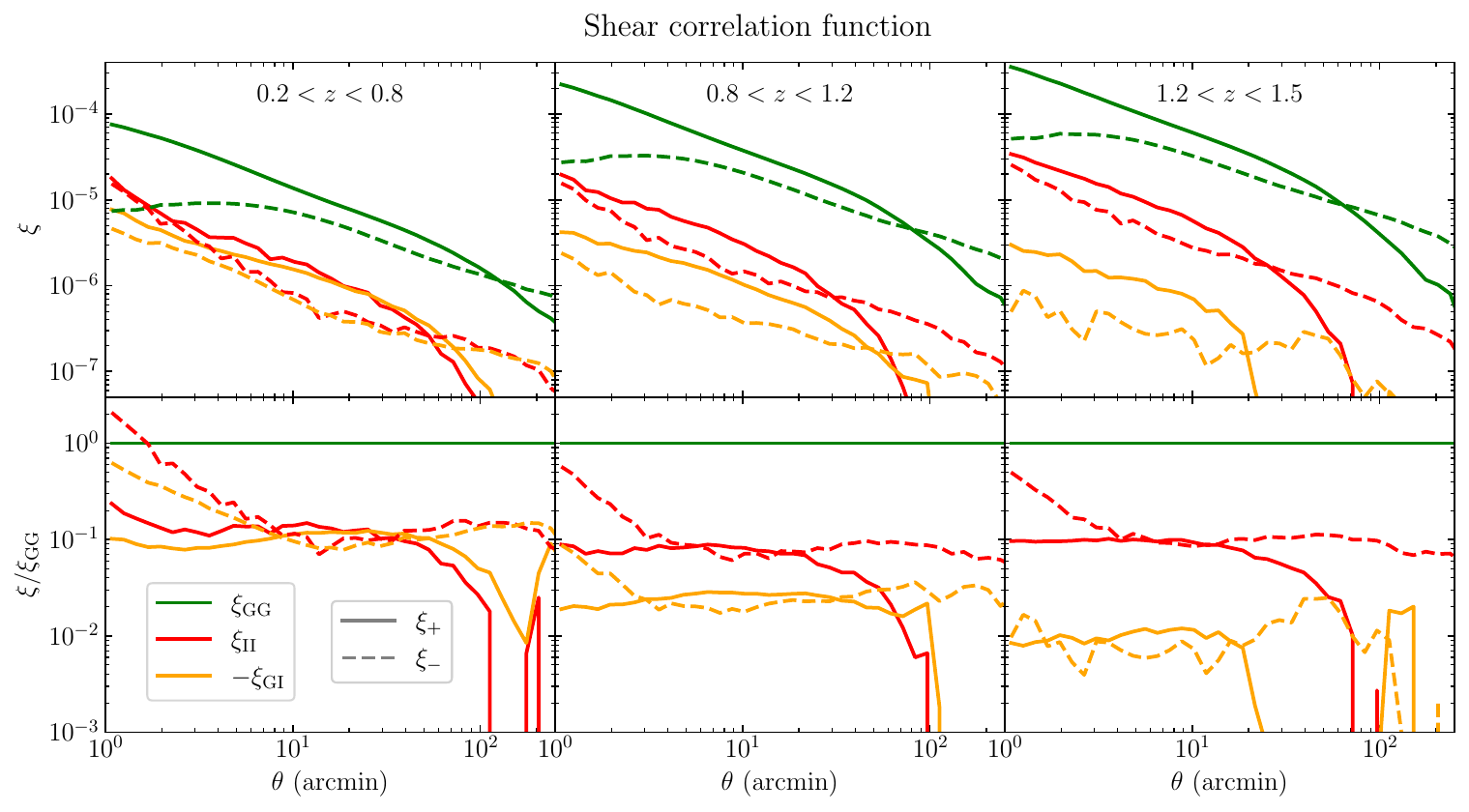}
    \caption{Shear auto-correlation of the pure weak lensing component $\xi_{\mathrm{GG}}$ (green lines), auto-correlation of the pure intrinsic alignment component $\xi_{\mathrm{II}}$ (red lines), and their cross-correlation $\xi_{\mathrm{GI}}$ with inverted sign (yellow lines). The solid and dashed lines indicate the $\xi_+$ and $\xi_-$ statistics, respectively. The top panels show the correlation function, while the bottom panels display the ratio with respect to the auto-correlation $\xi_{\mathrm{GG}}$. Each column refers to a different redshift bin. The II and GG signals grow with redshift while maintaining a roughly constant $\xi_{\mathrm{II}}/\xi_{\mathrm{GG}}$ ratio, whereas the GI signal decreases in relative amplitude, reflecting the redshift evolution of the IA signal.}
    \label{fig:corr_func_zbins}
\end{figure*}

To validate our pipeline, we begin by examining the shear correlation function and compare it with theoretical predictions, as shown in Figure~\ref{fig:corr_func_check}. We select galaxies from our catalogue that reside in each of our redshift bins, and compute the auto-correlation of the gravitational and intrinsic shear components, labeled ``GG'' and ``II'' respectively, as well as the cross-correlation function, labeled ``GI''. The GI term, as expected, exhibits a negative sign. This anti-correlation arises because the intrinsic shape of (foreground) galaxies tends to align radially with the large-scale gravitational potential, pointing toward overdense regions, while the lensing shear of (background) galaxies is tangential around the same overdense regions.

We produce theoretical predictions for the above quantities with the {\sc CCL}\footnote{https://github.com/LSSTDESC/CCL} library \citep{Chisari2019}. To do so, we start by computing the shear E-mode auto and cross angular power spectra according to equations (50), (51), and (52) of \citet{Lamman2024}. The corresponding formulae require the source redshift distribution as input (shown in Figure~\ref{fig:redshift_mass_ellipticity_dist}), as well as the 3D matter power spectrum in the redshift range of interest, which in this work is computed with the {\sc Halofit} formula \citep{Takahashi2012}. In the case of the II and GI power spectra, we employ the NLA model \citep{Bridle2007}, which assumes a linear proportionality between the intrinsic shape of galaxies and the local tidal gravitational field. Although similar to the LA model \citep{Catelan2001}, the NLA model also incorporates the nonlinear evolution of the matter density field. The proportionality is expressed in terms of the $A_1$ parameter, consistent with the definitions in equation (11) of \citet{Samuroff2023}. Once the angular power spectra are computed, these can be converted into correlation functions according to Equation~(\ref{eq:hankel_transform}).

Overall, we observe a qualitatively good agreement between our results and the theoretical predictions in the angular range we investigate. The NLA model provides a reasonably good fit to the data on scales where its assumptions hold: particularly at intermediate and large angular separations. Best-fit values of the alignment amplitude, obtained by minimizing the root mean square error between the NLA prediction and our data for $\theta \gtrsim 2$ arcmin, are approximately $A_1 \approx [2.06, 2.55, 3.52]$ for the low-, intermediate-, and high-redshift bin respectively.

Figure~\ref{fig:corr_func_check} shows that the NLA model agrees with our simulated pipeline to within $\approx50\%$. While such deviations are expected given the limitations of the NLA model, this large discrepancy serves as motivation for our fully non-linear simulated pipeline. This enables a more accurate characterization of the IA signal and gives us direct access to higher-order statistics measurements that include IA contributions without relying on simplifying assumptions.

\subsection{Redshift dependence}
\label{subsec:redshift_dep}

Having validated our basic results, we now move on to investigate the redshift dependence of the shear correlation function, as well as the convergence angular power spectrum, its PDF, and peaks and minima.

\subsubsection{Shear Correlation function}
\label{subsubsec:redshift_dep_gamma_corr}

In Figure~\ref{fig:corr_func_zbins}, we present the shear correlation functions computed for the three redshift bins investigated in this work. Looking at the gravitational (GG) and intrinsic (II) auto-correlations, we see that, as the redshift increases, both increase in amplitude, maintaining roughly the same ratio of $\xi_{\mathrm{II}}/\xi_{\mathrm{GG}}\approx0.1$ at intermediate angular scales; i.e.~$\theta \approx 10 \,\mathrm{arcmin}$, as shown by the red curve in the bottom sub-panels. The redshift evolution of $\xi_{\mathrm{II}}$ has been reported in previous numerical studies \citep[see e.g. D23;][]{Chisari2015, Zjupa2022}, and the increase of its  amplitude with redshift can be explained by the following two arguments. First, it is consistent with the theoretical framework in which a lower non-linear disruption of alignments at high redshift is expected \citep[see e.g.][]{Lamman2024}. Second, we recall that, as seen in Section~\ref{subsec:gal_cat_selection}, our three redshift bins are characterised by a decreasing comoving width with redshift. It is important to note that, as the width of the shell increases, regions that are progressively more separated in physical space, and thus less correlated, will be projected together, diluting the overall signal.

Regarding the gravitational-intrinsic (GI) cross-correlation function, we see that the amplitude remains relatively constant for different redshift bins, with the ratio $-\xi_{\mathrm{GI}}/\xi_{\mathrm{GG}}$ going from $\approx 0.1$, to $\approx 0.02$, to $\approx 0.008$, at low, intermediate and high redshifts respectively. Furthermore, it is interesting to note that the GI signal shows a decreasing trend relative to the II signal with increasing redshift. In particular, in the lowest redshift bin, we find that $\xi_{\mathrm{GI}}$ and $\xi_{\mathrm{II}}$ have similar amplitude and scale dependence, especially at angular scales larger than $\approx10$ arcmin. This redshift trend can help us understand the substantial redshift evolution of the impact of IA on WL convergence statistics, which we investigate in the next section.

\subsubsection{Convergence power spectrum}\label{subsubsec:redshift_dep_kappa_pk}

\begin{figure*}
    \centering
    \includegraphics[width=\textwidth]{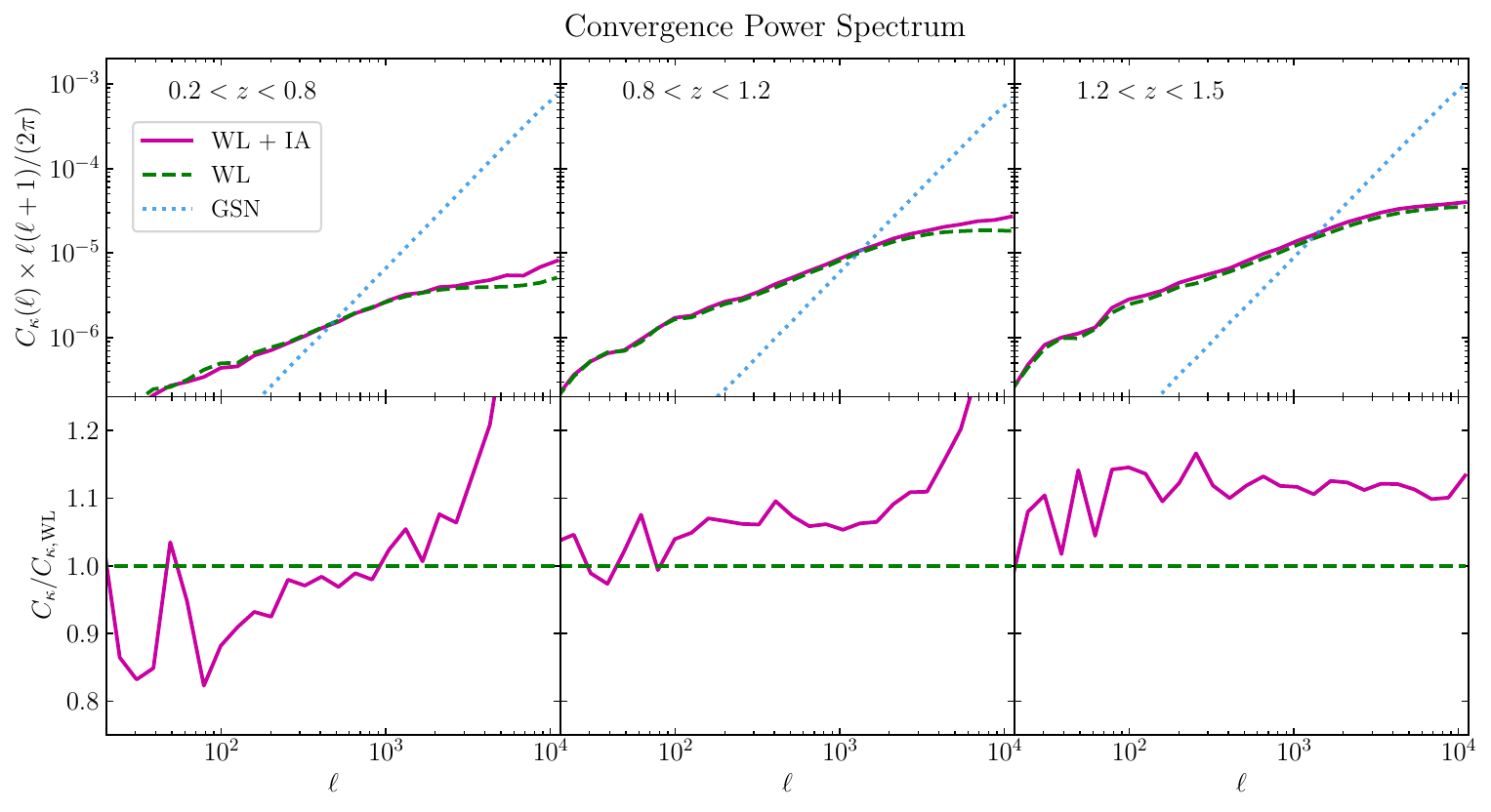}
    \includegraphics[width=\textwidth]{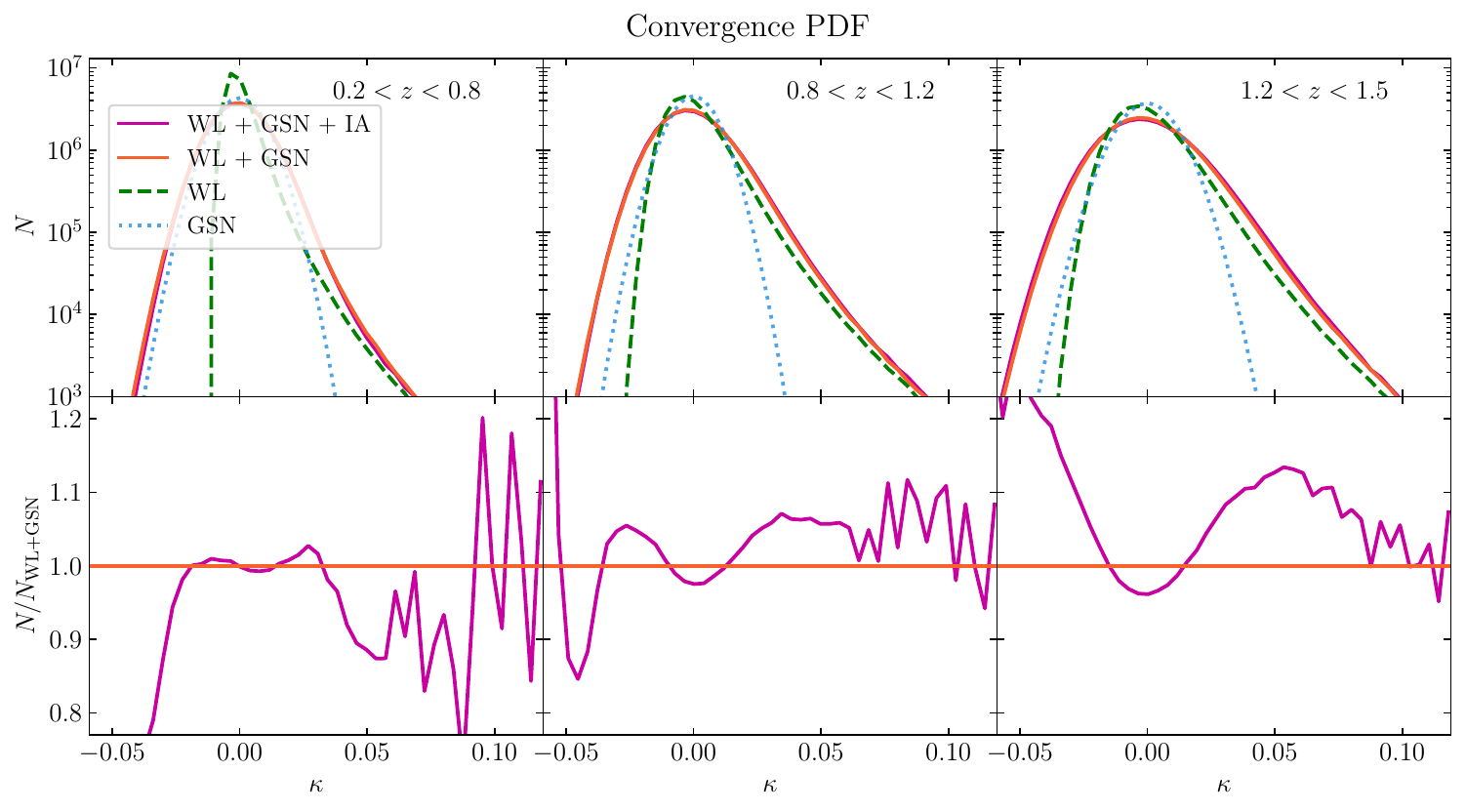}
    \caption{{\it Top:} Angular power spectrum of convergence maps. Dashed green lines show the pure weak lensing (WL) signal; light blue dotted lines indicate galaxy shape noise (GSN); purple solid lines represent maps including both WL and IA, with GSN subtracted. Bottom sub-panels show ratios relative to the WL case. Each column corresponds to a redshift bin. IA introduces a scale- and redshift-dependent modulation of the WL power spectrum, due to the interplay between negative GI and positive II terms causing suppression at low redshift and large scales, and up to $\sim20\%$ enhancement at small scales and high redshift.\\
    {\it Bottom:} PDF of convergence maps; purple solid lines correspond to maps including WL, GSN, and IA; orange solid lines show the same case but with randomized intrinsic orientations; dashed green lines show the pure WL signal; light blue dotted lines indicate the GSN. Bottom sub-panels show ratios with respect to the WL + GSN case. Maps were smoothed with a 2 arcmin Gaussian kernel before PDF computation. Each column corresponds to a redshift bin. IA alters the convergence PDF in a redshift-dependent way that can be partially explained by considering the power spectrum: boosted power at high redshift broadens the PDF, suppression at low redshift narrows it; at intermediate redshift, enhanced small-scale power induces non-Gaussian PDF distortions.
    }
    \label{fig:pk_pdf_zbins}
\end{figure*}

In the top panel of Figure~\ref{fig:pk_pdf_zbins}, we show the convergence power spectrum in the three redshift bins investigated in this work. After computing the convergence maps according to Section~\ref{subsec:gamma_to_kappa}, we measure the angular power spectrum for the pure WL signal (green dashed curve), and the GSN signal (blue dotted curve), using the corresponding convergence fields. To obtain the power spectrum of the WL convergence field with the IA signal (``WL + IA'', purple solid curve), i.e.~no GSN contribution, we first compute the power spectrum of the convergence field corresponding to WL + GSN + IA, and then subtract the GSN power spectrum. 

First, we observe that the GSN curve follows a white noise spectrum, as expected \citep[see e.g.][]{Kaiser1995}. We note that the noise component crosses the WL spectrum at $\ell \approx  500$ in the low redshift bin, and at $\ell \approx  1500$ in the two remaining redshift bins. Looking at the ratio between the WL + IA and WL curves, we see that the IA signal has a significant qualitative impact on the WL signal over all redshift bins. At high redshift, we observe a $\approx 10 \%$ power enhancement at all scales, which is slightly weaker at the largest angular scales plotted here. At the intermediate redshift, we see a stronger trend in $\ell$, which is weaker at large-intermediate scales, giving a $\approx 5 \%$ enhancement at $10^2 \lesssim \ell \lesssim  10^3$, but becoming progressively larger at the smaller angular scales, reaching $20\%$ at $\ell\approx5000$. Moving to the low redshift bin, we see a $\approx10\%$ suppression at large scales, then the suppression progressively decreases, with the ratio approaching one at intermediate angular scales ($\ell\approx900$), which then turns into an enhancement (similar to the intermediate redshift bin), reaching $20\%$ at $\ell\approx4000$. 

The strong redshift dependence observed in the power spectrum can be better understood by comparing it to the shear correlation function in Figure~\ref{fig:corr_func_zbins}. In particular, let us start by noting that the WL + IA power spectrum, being the sum of two components, i.e.~gravitational and intrinsic, can be expressed as $C_{\kappa, \mathrm{WL + IA}} = C_{\kappa, \mathrm{GG}} + 2C_{\kappa, \mathrm{GI}} + C_{\kappa, \mathrm{II}}$. Additionally, from Figure~\ref{fig:corr_func_zbins}, we note that the values of $|\xi_\mathrm{GI}|$ and $|\xi_\mathrm{II}|$ are roughly comparable, especially at scales larger than 10 arcmin. Finally, we recall that the gravitational and intrinsic shear fields are anti-correlated; i.e.~$\xi_\mathrm{GI}$ is negative (and so is $C_{\kappa, \mathrm{GI}}$). By combining the above information, it becomes clearer that the $C_{\kappa,\mathrm{WL+IA}} / C_{\kappa,\mathrm{WL}}$ ratio is less than one at large-to-intermediate scales in the low redshift bin, which can be explained in terms of the cross-correlation term (GI) dominating the intrinsic auto-correlation term (II) in this bin. On the other hand, the cross term becomes progressively less important at intermediate and high redshift, which explains the increasing values of $C_{\kappa,\mathrm{WL+IA}} / C_{\kappa,\mathrm{WL}}$ seen in these bins.

\subsubsection{Convergence PDF, peaks and minima}
\label{subsubsec:redshift_dep_kappa_pdf_peaks_minima}

\begin{figure*}
    \centering
    \includegraphics[width=\textwidth]{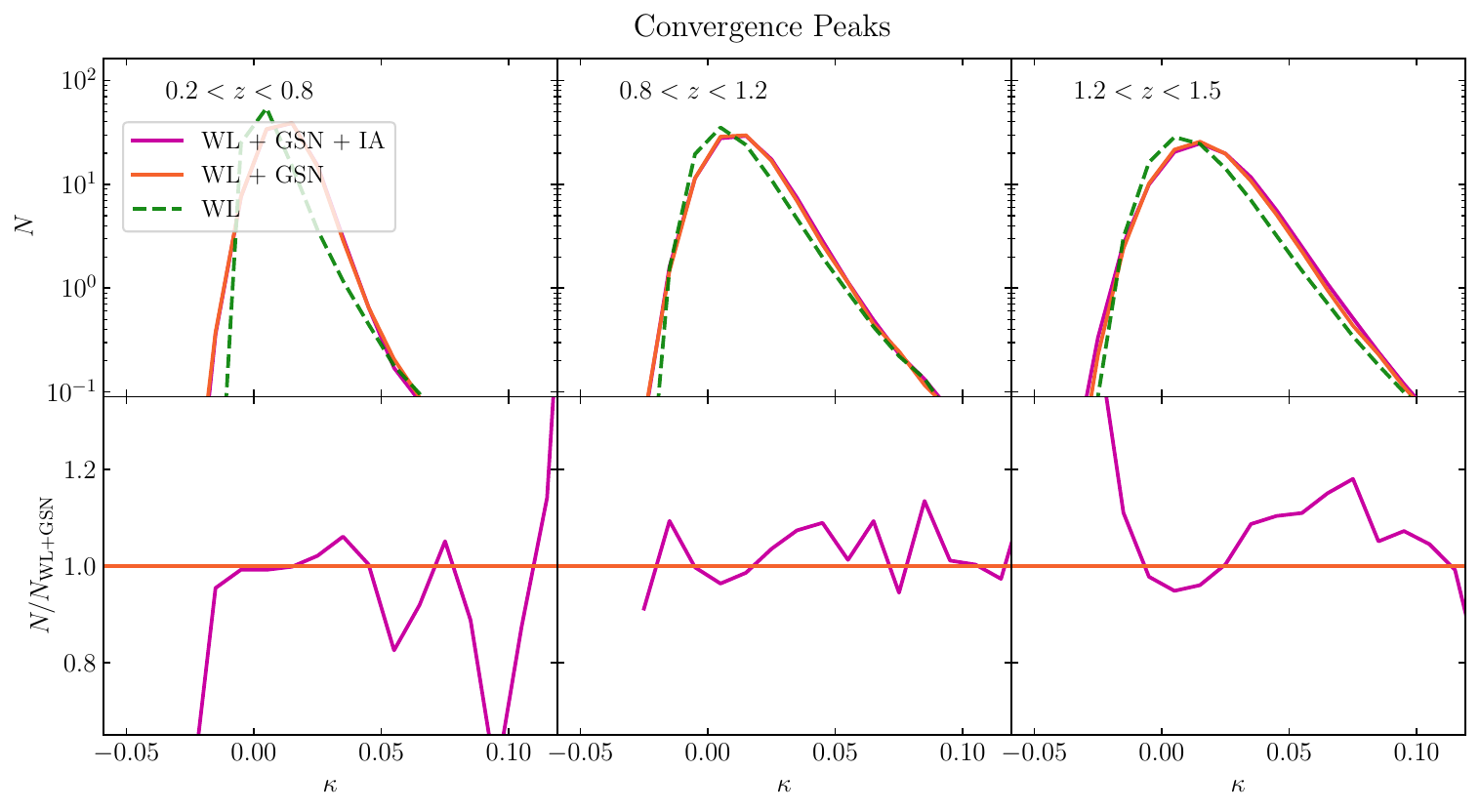}
    \includegraphics[width=\textwidth]{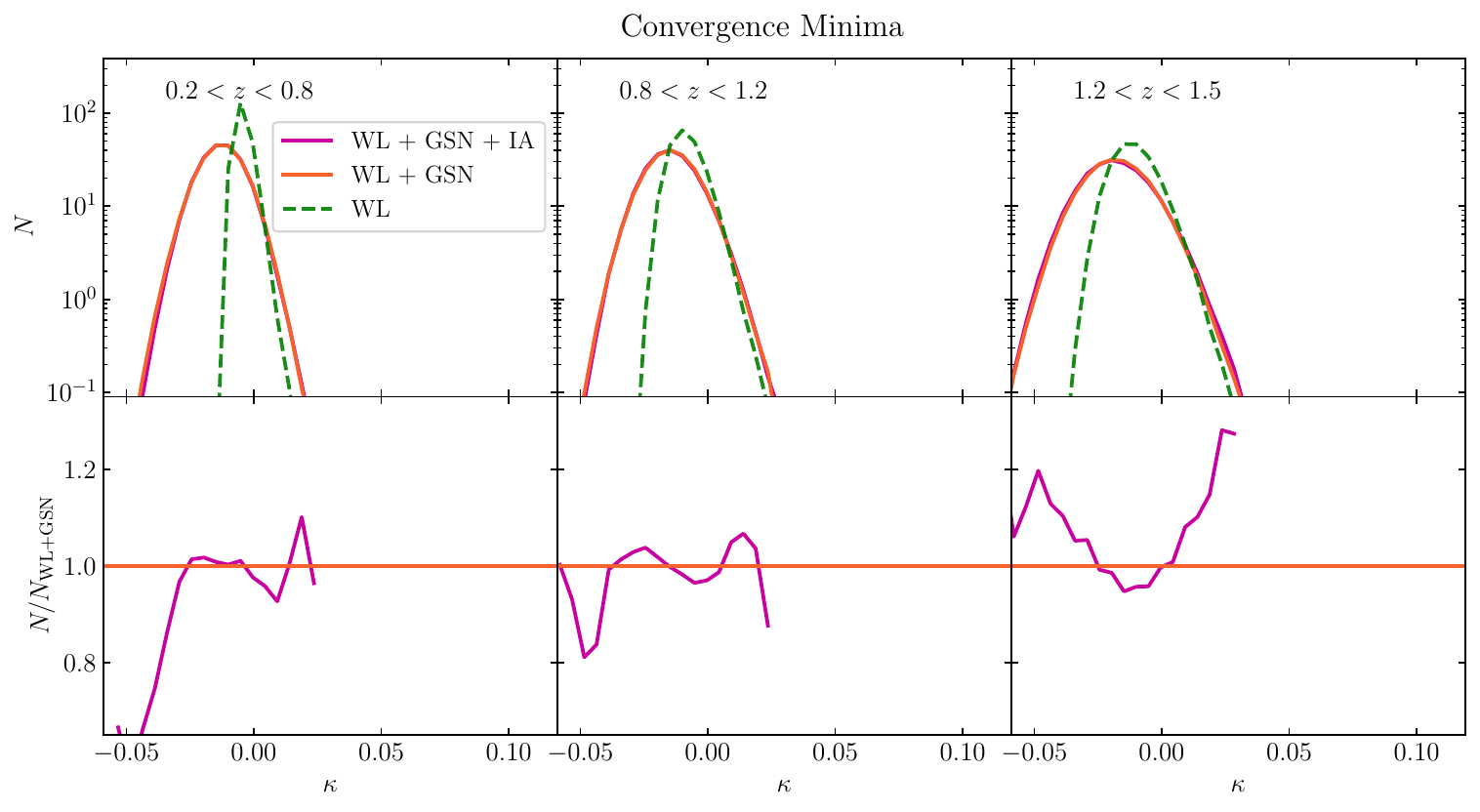}
    \caption{{\it Top (bottom):} Peak (minimum) counts distribution of convergence maps; the purple solid lines refer to maps containing the total signal comprehensive of WL, intrinsic galaxy shapes (GSN), and orientations (IA); the orange solid lines refer to a version of the previous case where intrinsic ellipticities have been preserved, but intrinsic orientations have been randomized; the dashed green lines refer to the pure weak lensing signal. The maps have been smoothed with a 2 arcmin Gaussian kernel before the computation of the statistic. The top sub-panels show the peak (minimum) counts, while the bottom sub-panels give the ratio with respect to the WL + GSN case. Each column refers to a different redshift bin. The impact of IA on peak and minimum counts mirrors the redshift-dependent trends observed in the convergence PDF (shown in Figure~\ref{fig:pk_pdf_zbins}), with tail suppression at low redshift, enhancement at high redshift, and more complex, small-scale-driven distortions at intermediate redshift.}\label{fig:peaks_minima_zbins}
\end{figure*}

In the bottom panel of Figure~\ref{fig:pk_pdf_zbins}, we show the convergence PDF for the three redshift bins investigated in this work. First, we note that the PDF of the pure WL convergence field (green dashed curve) shows a significant positive skewness, which arises from nonlinear structure growth \citep[see e.g.][]{Valageas2000}. In contrast to this, we see that the pure GSN convergence field (blue dotted curve) follows a Gaussian distribution, consistent with white noise. The convergence field that contains both of the components described above, i.e. WL + GSN (orange solid line), is characterised by a PDF that results from the convolution of the PDFs of its individual components. Therefore, as we can see in the top sub-panels, the PDF of the WL + GSN convergence field is broader, less skewed, and has diluted non-Gaussian features with respect to the pure WL convergence PDF.

Let us now investigate the impact of IA on the convergence PDF. To do so we consider the lower sub-panels, where the ratio between the WL + GSN + IA and WL + GSN cases is shown. To ease the interpretation, we now discuss the redshift bins in an order where the underpinning physics of interest increases in complexity, which is high-$z$, then low-$z$, followed by intermediate-$z$. For the high-$z$ bin, we see that the impact of IA leads to a clear broadening of the PDF. This is characterised by an enhancement of the high- and low-$\kappa$ tail that reaches $\approx20\%$ and $\approx13\%$ at $\kappa\approx-0.04$ and $\kappa\approx0.055$ respectively; and a suppression of the central region, that reaches $\approx4.5\%$ at $\kappa\approx0$. To better understand the nature of this broadening, let us notice that, as seen in the previous section, the presence of IA boosts the convergence power spectrum at all scales in this redshift bin. Therefore, a boost in power will increase the field's variance, resulting in a broader PDF, since the width of the PDF is set by the variance of the field. Moving our attention to the low-redshift bin, and using the same argument given above, the reduction in power leads to a narrowing of the convergence PDF when IA is included. This appears as a suppression of the high- and low-$\kappa$ tails that reaches $\approx20\%$ and $\approx12\%$ at $\kappa\approx-0.03$ and $\kappa\approx0.055$ respectively, with a distortion of the central region that remains below $\approx3\%$. Focusing now on the intermediate redshift bin, we observe that IA distorts the PDF in a more complex way, showing intermediate features between the other two redshift bins. Given that the power is enhanced the most at small scales with the inclusion of IA in this redshift bin, we can conclude that the impact from IA that we observe on the PDF is driven by the small scales. Noting that these scales are non-linear, this leads to modifications of the higher-order moments of the convergence field, which manifest as more complex alterations to the shape of the PDF, namely the multiple crossings of $N_{\rm{WL+GSN+IA}}/N_{\rm{WL+GSN}}=1$ at low $\kappa$.  The impact in this bin is characterised by a suppression of the low-$\kappa$ tail that reaches $\approx15\%$ at $\kappa\approx-0.045$, and distortions in the rest of the PDF that reach $\approx12\%$ at $\kappa\approx0.08$.

We continue our analysis of the redshift dependent impact of IA on WL convergence statistics by studying peak and minimum counts, shown respectively in the upper and lower panels of Figure~\ref{fig:peaks_minima_zbins}. Looking at the overall picture, we first remind the reader that distortions in the peaks and minima distributions are closely linked to distortions in the convergence PDF, particularly in its tails. Indeed, it can be clearly seen that the impact of IA on both peaks and minima, at each redshift bin, show trends that are qualitatively consistent with the impact on the PDF. We also note that, while the trends are similar to the PDF, the peaks and minima statistics are measured with coarser bins relative to the PDF, as there are far fewer extrema than pixels in a given WL map, which leads to the coarser ratios presented in Figure~\ref{fig:peaks_minima_zbins} compared to the PDF.

In the case of the peak counts, focusing on the low-redshift bin, we observe that the inclusion of IA leads to a suppression of the tails that reaches $\approx20\%$ at $\kappa\approx-0.019$, and $\approx18\%$ at $\kappa\approx0.055$. Regarding the intermediate redshift bin, we observe that the distribution tends to fluctuate, similar to the PDF, showing an alternation of enhancements and suppressions that never exceed $\approx13\%$. In the case of the high-redshift bin, we observe an enhancement of the tails that reaches $\approx30\%$ at $\kappa\approx-0.02$, and $\approx17\%$ at $\kappa\approx0.075$.

Finally, in the case of the minimum counts, looking at the low-redshift bin we find a suppression in the low-$\kappa$ tail that reaches $\approx30\%$ at $\kappa\approx0.041$, and fluctuations in the high-$\kappa$ tail that reach $\approx10\%$ at $\kappa\approx0.018$. In the intermediate redshift bin, we observe fluctuations below $\approx7\%$ over the range $-0.04\lesssim\kappa\lesssim0.02$. In the case of the high-redshift bin we observe an enhancement of the tails that reaches $\approx20\%$ at $\kappa\approx-0.05$, and $\kappa\approx0.02$.

Before moving on to the next section, we remark that our findings are consistent with a very recent study by \citet{Lee2025}, who performed a similar analysis based on the {\sc IllustrisTNG} project, finding IA to impact the convergence power spectrum, PDF, peaks and minima with comparable magnitude to what is detected in our study. However, we point out that a direct comparison with this study is not possible, since, amongst other things, their analysis considered the total source population without redshift binning and uses different selection criteria to build the catalogue (e.g. minimum number of star particles). Moreover, our analysis offers improved robustness and sampling of large scale modes as it is based on a box volume that is $\approx15$ times bigger.

\subsection{Galaxy stellar mass dependence}
\label{subsec:mass_dependence}

\begin{figure*}
    \centering
    \includegraphics[width=\textwidth]{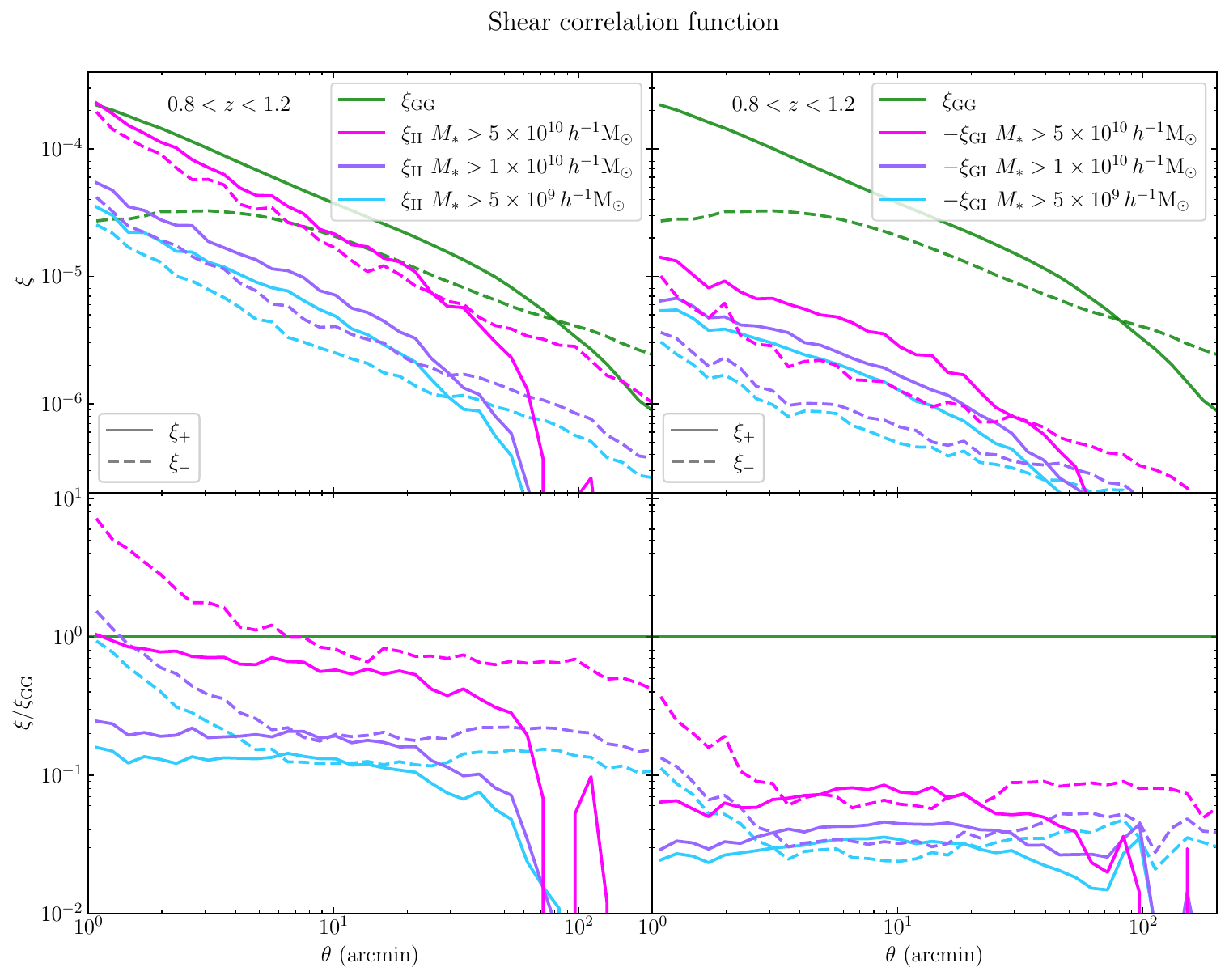}
    \caption{{\it Left (right)}: the light blue, violet and pink curves refer to the auto-correlation of the pure intrinsic alignment component $\xi_{\mathrm{II}}$ (cross-correlation between intrinsic and gravitational shear $\xi_{\mathrm{GI}}$) for increasingly higher minimum stellar mass thresholds. The green lines refer to the auto-correlation of the gravitational shear $\xi_{\mathrm{GG}}$. The top panels give the correlation function, while the bottom panels display the ratio with respect to the auto-correlation $\xi_{\mathrm{GG}}$. The shear intrinsic auto- and gravitational-intrinsic cross-correlation amplitudes increase with stellar mass, with a stronger effect on $\xi_{\mathrm{II}}$ than on $\xi_{\mathrm{GI}}$. The case with the highest minimum stellar mass cut shows IA signals comparable to the WL signal, especially at small angular scales.}
    \label{fig:corr_func_massbins}
\end{figure*}

Having investigated the redshift dependence of the impact of IA on a variety of WL statistics, we conclude our analysis by studying the galaxy stellar mass dependence of the shear intrinsic autocorrelation and the gravitational-intrinsic cross-correlation. To do so, we first select three possible minimum stellar mass thresholds $M_\mathrm{min}=[5\times10^9, 1\times10^{10}, 5\times10^{10}] \, \Msh$ (indicated by the grey dashed lines in the central panel of Figure~\ref{fig:redshift_mass_ellipticity_dist}) and apply them to our catalogue. Successively, we compute $\xi_{\mathrm{II}}$ and $\xi_{\mathrm{GI}}$ in our central redshift bin and present it in Figure~\ref{fig:corr_func_massbins}, where the left and right sub-panels refer to $\xi_{\mathrm{II}}$ and $\xi_{\mathrm{GI}}$, and increasing minimum stellar mass cuts are represented by the light blue, violet, and pink curves, respectively. For a direct comparison with the WL signal, we also include $\xi_{\mathrm{GG}}$ in all panels, represented by the green curves. 

By looking at $\xi_{\mathrm{II}}$ and $\xi_{\mathrm{GI}}$, we notice that both quantities experience a nearly scale-independent increase in amplitude with increasing stellar mass thresholds. To quantify such an increase in the theoretical context, we have performed a fit of the NLA model to our results, finding best matches with $A_1=[3.31, 4.07, 7.93]$ for $M_\mathrm{min} = [5\times10^9, 1\times10^{10}, 5\times10^{10}] \, \Msh$. This trend is qualitatively consistent with observations \citep[see e.g.][]{Joachimi2013, Singh2015, Samuroff2023} and previous simulation results \citep[see e.g. D23;][]{Chisari2015, Hilbert2017, Zjupa2022}, indicating that more massive galaxies, which are often of early-type and reside in denser environments, are more coherently aligned with the surrounding large-scale structure, thereby amplifying the intrinsic alignment signals.

Focusing on the lower sub-panels, it becomes clearer that the mass dependence has a stronger impact on the amplitude of $\xi_{\mathrm{II}}$ than on that of $\xi_{\mathrm{GI}}$. This behavior can be directly explained in the context of the NLA model, by noting that it predicts $\xi_{\mathrm{II}} \propto A_1^2$ and $\xi_{\mathrm{GI}} \propto A_1$. We observe that, considering the lowest and highest mass cuts, and looking at intermediate angular scales, $\xi_{\mathrm{II}}/\xi_{\mathrm{GG}}$ changes from $\approx0.12$ to $\approx0.65$, while $-\xi_{\mathrm{GI}}/\xi_{\mathrm{GG}}$ goes from $\approx0.03$ to $\approx0.07$. Notably, we find that a mass cut of $5\times10^{10} \Msh$ boosts the amplitude of $\xi_{\mathrm{II}}$ to a level that is of the same order of magnitude as $\xi_{\mathrm{GG}}$. Interestingly, we see that $\xi_{-,\mathrm{II}}$ dominates $\xi_{-,\mathrm{GG}}$ at all angular scales smaller than $\approx7$ arcmin, with $\xi_{-,\mathrm{II}}/\xi_{-,\mathrm{GG}}$ reaching $170\%$ at the arcmin scale.

The strong stellar mass dependence of the IA signal highlighted above has important implications for interpreting our results throughout the paper. As stated in the previous paragraph, a higher mass cut will boost $\xi_{\mathrm{II}}$ quadratically and $\xi_{\mathrm{GI}}$ linearly, leading the II component to increasingly dominate over the GI component. In the light of the discussions in Section~\ref{subsec:redshift_dep}, we expect the effects of higher mass cuts on convergence statistics to be qualitatively similar to the ones observed in the high-$z$ bin of this study, where the II component dominates over the GI component the most.


\section{Conclusions and outlook}
\label{sec:conclusions}

In this paper, we have presented a fully non-linear and self-consistent forward model to study the impact of intrinsic alignments (IA) on weak lensing (WL), based on the flagship $740\,\textrm{Mpc}$ full-hydro simulation of the MillenniumTNG project. Starting from the particle lightcone covering an octant of the sky in the redshift range $z=[0, 1.5]$, we  have identified galaxies with the {\sc Subfind} halo and galaxy finder, and we simultaneously computed both the intrinsic and gravitationally induced (extrinsic) shear signals for each galaxy. The intrinsic shear component is computed from the luminosity-weighted inertia tensor of stellar particles. The extrinsic (gravitational) shear is extracted via full-sky ray-tracing on the same simulation using the {\small DORIAN} code (i.e., without invoking the Born approximation). This allowed us to directly evaluate the impact of IA on a range of WL observables, including shear correlation functions, convergence power spectra, and higher-order statistics such as the convergence PDF, peak, and minimum counts. All of these quantities where studied for three different redshift bins, with edges at $z = [0.2, 0.8, 1.2, 1.5]$.

To validate our pipeline, we compared the measured shear correlation functions with theoretical predictions derived using the Nonlinear Alignment model (NLA) for IA. We found good qualitative agreement across the redshift bins and angular scales considered, with the best fit obtained for an alignment amplitude of $A_1 \approx [2.06, 2.55, 3.52]$ for the low-, intermediate-, and high-redshift bin respectively. 

We analysed the redshift evolution of the shear correlation functions and found that both the gravitational and intrinsic auto-correlations ($\xi_{\mathrm{GG}}$ and $\xi_{\mathrm{II}}$ respectively) increase in amplitude with redshift, while maintaining a roughly constant ratio of $\xi_{\mathrm{II}}/\xi_{\mathrm{GG}}\approx0.1$ at intermediate angular scales ($\theta\approx10\,\mathrm{arcmin}$). This redshift evolution of $\xi_{\mathrm{II}}$ is consistent with reduced non-linear disruption of alignments at high redshift and the narrower comoving width of the redshift bins, which reduces signal dilution. In contrast, the gravitational-intrinsic cross-correlation $\xi_{\mathrm{GI}}$ shows a strong decline in relative amplitude, with $-\xi_{\mathrm{GI}}/\xi_{\mathrm{GG}}$ decreasing from $\approx0.1$ to $\approx0.008$ from low to high redshift, indicating a diminishing role of the GI term at higher redshift.

We then focused on WL convergence statistics and computed these both in the presence and absence of IA. In the case of the convergence power spectrum, we found that the IA signal introduces a notable modification to the WL power spectrum, with redshift- and scale-dependent behavior: a $\sim10\%$ enhancement at high redshift, a scale-dependent increase peaking at $\sim20\%$ at small scales at intermediate redshift, and a transition from $\sim10\%$ suppression at large scales to $\sim20\%$ enhancement at small scales at low redshift.  This complex trend is attributed to the interplay between the gravitational-intrinsic (GI) and intrinsic-intrinsic (II) components, where the GI term dominates at low redshift, leading to suppression, while its influence diminishes with redshift, allowing the II contribution to enhance the signal at higher redshifts.

We examined the redshift-dependent impact of IA on the convergence PDF, peak, and minimum statistics. At high redshift, IA induces a broadening of the PDF, with tail enhancements up to $\approx20\%$ and $\approx13\%$ at $\kappa \approx -0.04$ and $\kappa \approx 0.055$, respectively, and a central suppression of $\approx4.5\%$, consistent with increased power on small scales. Conversely, at low redshift, the reduced power leads to a narrowing of the PDF, with tail suppressions of $\approx20\%$ and $\approx12\%$. These distortions propagate to peak and minimum counts, which mirror the PDF trends: in the low-redshift bin, peak and minimum tails are suppressed by up to $\approx20$--$30\%$, while in the high-redshift bin, enhancements reach $\approx30\%$ for peaks and $\approx20\%$ for minima; intermediate redshift bins show alternating enhancements and suppressions below $\approx13\%$.

We further examined the dependence of IA on galaxy stellar mass by applying three increasing minimum stellar mass thresholds and computing the corresponding shear intrinsic and cross-correlations in the central redshift bin. We find that both $\xi_{\mathrm{II}}$ and $\xi_{\mathrm{GI}}$ increase nearly uniformly across scales with stellar mass, consistent with the expectation that more massive galaxies exhibit stronger alignment with the large-scale structure. Notably, $\xi_{\mathrm{II}}$ shows a stronger sensitivity to mass, with its amplitude reaching levels comparable to $\xi_{\mathrm{GG}}$ for the most massive galaxies, and even dominating $\xi_{-,\mathrm{GG}}$ at scales below $\approx7$ arcminutes.

The results obtained here can also be viewed as an illustration of the power of the new methodology we have explored in this study, namely to study a seamless lightcone obtained directly from a large-volume cosmological hydrodynamical simulation of galaxy formation. This approach is free of many approximations made in more conventional analysis and thus offers the prospect of improved quantitative accuracy and reliability.
Looking ahead, future extensions of this work could involve a more detailed exploration of the IA signal and its impact on WL as a function of galaxy morphology, colour, or environment, as well as the influence of different shape estimators on the measured alignment. Furthermore, the framework developed in this work can be used to explore the impact of intrinsic alignments on other observables, including galaxy-galaxy lensing. The galaxy catalogue and associated shear measurements produced in this work will be soon publicly released, providing a valuable resource for a wider range of cosmological and astrophysical studies.

\section*{Acknowledgements}

FF would like to thank Max E. Lee, Fabian Schmidt, Martin Reinecke, Soumya Shreeram, Nikolina Šarčević, Claire Lamman, and Ken Osato, for interesting and useful discussions. SB is supported by the UK Research and Innovation (UKRI) Future Leaders Fellowship [grant number MR/V023381/1].  VS and LH acknowledge support from the Simons Foundation through the "Learning the Universe" project.

\section*{Data Availability}

The galaxy shear catalogue from this work will be publicly released in the near future as part of the MTNG data release. In addition to shear quantities, the catalogue will include a range of galaxy properties such as stellar mass, size, star formation rate, and luminosities in multiple photometric bands, making it suitable for a wide range of astrophysical applications. The ray-tracing code {\small DORIAN} is publicly available on GitLab\footnote{\url{https://gitlab.mpcdf.mpg.de/fferlito/dorian}}. The data underlying this article will be shared upon reasonable request to the
corresponding authors.

\bibliographystyle{mnras}
\bibliography{main}

\bsp
\label{lastpage}
\end{document}